\documentclass{JHEP3}
\pdfoutput=1
\usepackage{amssymb, amsmath, bbm}
\usepackage{graphicx}

\def \ni {\noindent}

\def \mc {\mathcal}
\def \mbb {\mathbb}
\newcommand\cN{\mathcal{N}}

\title{Kinetic mixing of U(1)s for local string models}
\author{Mathew Bullimore, Joseph P. Conlon, Lukas T. Witkowski \\
 Rudolf Peierls Centre for Theoretical Physics, 1 Keble Road, Oxford OX1 3NP, UK \\

 E-mail: \email{m.bullimore1@physics.ox.ac.uk},
  \email{j.conlon1@physics.ox.ac.uk}, \email{l.witkowski1@physics.ox.ac.uk}
}

\vspace{1.5cm}

\preprint{OUTP-10/22P}

\abstract{We study kinetic mixing between massless U(1)s in toroidal orbifolds with D3-branes at orbifold singularities.
We focus in particular on $\mbb{C}^3/\mbb{Z}_4$ singularities but also study $\mbb{C}^3/\mbb{Z}_6$ and $\mbb{C}^3/\mbb{Z}_6^{'}$
singularities.
We find kinetic mixing can be present and describe the conditions for it to occur. Kinetic mixing comes from winding modes
in the $\mc{N}=2$ sector of the orbifold.
 If kinetic mixing is present its size depends only on the complex structure modulus of the torus and is independent of the
 K\"ahler moduli.
We also study gauge threshold corrections for local $\mathbb{Z}_M \times \mathbb{Z}_N$ orbifold models finding
that, consistent with previous studies, gauge couplings run from the bulk winding scale rather than the string scale.}

\begin{document}
\section{Introduction}
In (semi-)realistic string models there are often a
multitude of U(1) factors present. This is particularly true for models including D-branes (for a review of such models see
\cite{Blumenhagen}).
 Although some of these U(1)s become massive and decouple, there are no theoretical obstructions to having
 multiple massless U(1)s present in a consistent string model.
  These U(1)s can inhabit both the visible and hidden sectors of the model.
This is clearest in the context of local models involving branes at singularities, where gauge groups can be geometrically
separated in the compact space.
   Given the presence of several massless U(1)s in a theory,
    it is a natural question to ask whether kinetic mixing can occur in such a setup.
    This possibility can also be phenomenologically significant
    if one of the U(1)s participating in kinetic mixing is the weak hypercharge of the Standard Model.

In a theory with two U(1) factors the low-energy effective Langrangian density can contain the following terms
\begin{equation}
\mathcal{L} \supset -\frac{1}{4 g_a^2} F_{\mu \nu}^{(a)} F_{(a)}^{\mu \nu} -\frac{1}{4 g_b^2} F_{\mu \nu}^{(b)} F_{(b)}^{\mu \nu} +\frac{\chi_{ab}}{2 g_a g_b} F_{\mu \nu}^{(a)} F_{(b)}^{\mu \nu} + m_{ab}^2 A_\mu^{(a)} A_{(b)}^{\mu}.
\label{eq1}
\end{equation}
The kinetic mixing term is a renormalisable operator and appears with the parameter $\chi_{ab}$ which can be generated at an arbitrarily high energy scale. We will be considering the effect of string scale physics on the terms in (\ref{eq1}) and will only be interested
in cases where string scale contributions to $m_{ab}^2$ are vanishing.
The main focus of this work will be to calculate $\chi$ in a consistent string model.

The phenomenological interest in kinetic mixing is twofold. If hypercharge mixed kinetically with another massless U(1) from a hidden sector we could expect the existence either of millicharged particles carrying small amounts of electric charge or alternatively Standard Model
particles that are millicharged under exotic U(1)s \cite{HoldomMix, HoldomMilli, Durham2, 08054037}. This scenario is already the subject of recent experimental activity \cite{Laser,ALPS}. U(1)s with weak couplings to Standard Model particles have also been of interest for models of
dark matter explaining excess positron production in the galaxy \cite{Boehm}, and the particle physics
phenomenology of such weakly coupled U(1)s has been explored in \cite{Fayet1,Fayet2} and many subsequent works.

Kinetic mixing in string theory has been studied both in heterotic and type II string theories \cite{JMR,9710441,Durham,KM,0504232,09090017,09095440,09124206,10021840,09090515}.
The calculations of kinetic mixing presented so far were performed in type II either using CFT vertex operator approaches \cite{Durham,KM} or by
working in the effective SUGRA field theory \cite{Durham}. The magnitude of kinetic mixing effects in heterotic string theory was studied in \cite{JMR,0504232}. Our approach is complementary to those above, as it is performed for supersymmetric local D3 brane models
and is technically different, being performed using the background field formalism.
We will not only construct models and compute the kinetic mixing parameter, we will also describe
 general conditions to obtain kinetic mixing in toroidal orbifolds.

This paper is organised as follows. In section \ref{methods} we describe the methods needed to perform the string calculation of kinetic mixing, followed by section \ref{consistent} where we summarize how to construct consistent string models with D3-branes at orbifold singularities. In the remainder of the paper we construct models based on various orbifold singularities and examine whether kinetic mixing occurs. The $\mathbb{Z}_4$ orbifold (section \ref{kinmixD3}) is studied as the canonical example and its treatment contains the most detail. Finally,
in section \ref{general} we present general observations regarding kinetic mixing in toroidal orbifolds  illustrated with further examples. As the calculation of kinetic mixing is closely related to the study of gauge threshold corrections we also explore thresholds for local $\mathbb{Z}_M \times \mathbb{Z}_N$ orbifold models in the appendix \ref{ZMZN}.


\section{Methods for calculating kinetic mixing in string theory}
\label{methods}

Kinetic mixing is a 1-loop contribution to the interactions of gauge bosons. As such
the study of kinetic mixing in string theory is closely related to the calculation of gauge threshold corrections.
Our calculation of kinetic mixing will be based on the background field method that has been used to compute gauge threshold corrections in
\cite{B1,B2,B3,B4,ABD,0404087,C,CP,Honecker}. This approach differs from the vertex operator ansatz used in \cite{Durham}.

The background field method involves turning on a background spacetime magnetic field $F_{23}$ along the U(1) generators whose mixing we wish to examine:
\begin{equation}
F_{23}=B_1 Q^a + B_2 Q^b .
\end{equation}

The mixing parameter can then be extracted from the one-loop vacuum string amplitude $\Lambda$ which consists of contributions from torus, Klein bottle, M\"{o}bius strip and annulus worldsheet diagrams. As only open strings couple to the magnetic field we
 only ever need consider M\"{o}bius strip and annulus diagrams. Further, we will be working with models that are free of O-planes and hence
 we can exclusively consider the annulus diagram. The vacuum amplitude can be expanded in terms of the magnetic fields $B_1$ and $B_2$:
\begin{equation}
\Lambda= \Lambda_0 + \frac{1}{2} {\left(\frac{B_1}{2 \pi^2} \right)}^2 \Lambda_2^a + \frac{1}{2} {\left(\frac{B_2}{2 \pi^2} \right)}^2 \Lambda_2^b + \frac{B_1 B_2}{{\left(2 \pi^2 \right)}^2} \Lambda_2^{ab} + \dots
\end{equation}
where $\Lambda_0$ vanishes in a supersymmetric compactification. The $\mathcal{O}(B^2)$ contributions to the vacuum amplitude are finite for a consistent string model and the terms proportional to $B_1^2$ and $B_2^2$ contain information about
gauge threshold corrections. We will direct our attention towards $\Lambda_2^{ab}$ as the kinetic mixing parameter $\chi$ can be extracted from this term:\footnote{In the following we will not be careful and drop the indices on the magnetic fields $B_1$ and $B_2$. When speaking of the $\mathcal{O}(B^2)$-term in the context of kinetic mixing we will in fact only refer to the part that involves ${\Lambda}_2^{ab}$.}
\begin{equation}
{\left. \frac{\chi}{g_a g_b} \right|}_{\textrm{1-loop}} = \frac{1}{{4\pi}^2} {\Lambda}_2^{ab} .
\end{equation}
The contributions to kinetic mixing can be disassembled into the parts coming from massless and massive strings as has been shown in \cite{JMR}:
\begin{equation}
{\left. \frac{\chi}{g_a g_b} \right|}_{1-loop} (\mu)= \frac{1}{16 {\pi}^2} \ b_{ab} \ \int_{1/M_X^2}^{1/\mu^2} \frac{\textrm{d}t}{t} + \frac{1}{16 {\pi}^2} \ \int_{0}^{\infty} \frac{\textrm{d}t}{t} \Delta_{ab}(t),
\end{equation}
where $\chi$ is taken to be zero at the scale $M_X$ and the integral is over the modular parameter $t$ of the annulus. Massless strings contribute to $b_{ab}$ and massive strings give the term $\Delta_{ab}$. As we are mainly interested in U(1)s that are hidden from one another we will be exploring situations where $b_{ab}$ is zero whereas $\Delta_{ab}$ is non-vanishing. To get finite answers we will need to ensure that $\Delta_{ab} \rightarrow 0$ both for $t \rightarrow 0$ and $t \rightarrow \infty$. $\Delta_{ab}(t=0) =0$ comes from tadpole cancellation and
$\Delta_{ab}(t=\infty)=0$ comes from the fact that the $U(1)$s are hidden.

Kinetic mixing between hidden massless U(1)s thus derives from the $t$-dependence of massive string states. To calculate this in an orbifold setting we need to consider a global model as we will need to include strings that stretch across and wrap around the compactified dimensions. We will construct models based on toroidal orbifolds $\mathbb{T}^6/\mathbb{Z}_N$ where we place supersymmetric D3-branes at the orbifold singularities. The vacuum annulus amplitude in this context can be calculated as
\begin{equation}
\mathcal{A} = \int_0^{\infty} \frac{\textrm{d}t}{2t} \ \textrm{STr} \left(\frac{1+{\theta} + {\theta}^2 + \dots + {\theta}^{N-1}}{N} \frac{1+{(-1)}^F}{2} q^{(p^{\mu} p_{\mu} + m^2)/2} \right)
\end{equation}
where $q=e^{-\pi t}$, $\textrm{STr} = \sum_{bosons} - \sum_{fermion} \equiv \sum_{NS} - \sum_{R}$ and $\alpha^{\prime} = 1/2$. The supertrace is over string states that survive the GSO and the orbifold projections. In the following we will state the relevant vacuum amplitudes in a background magnetic field which we will use to explore kinetic mixing. However first we will review the physics of D3-branes at orbifold singularities.


\subsection{Orbifold singularities and the resulting spectrum}
We consider models where the orbifold twist ${\mathbb{Z}}_N$ acts crystallographically on the compact space $\mathbb{T}^6$ which factorises into three two-tori. The orbifold action on these tori is given by $\theta: z_i \rightarrow \exp (2 \pi i {\theta}_i) z_i, \ i=1,2,3$ which is identical with the geometric action on the complex scalars in the spectrum. The orbifold group is spanned by the elements ${\theta}^k$ where the exponent denotes $k$ applications of $\theta$.  In addition, we choose the orbifold to preserve $\mathcal{N}=1$ supersymmetry which enforces $\sum_{i=1}^3 {\theta}_i = 0$ mod 1.

The orbifold twist also has an effect on the Chan-Paton (CP) degrees of freedom. On each stack of coincident branes we choose an embedding of the form:
\begin{equation}
{\gamma}_{{\theta}^k} = \textrm{diag}( {\mathbbm{1}}_{n_0}, e^{\frac{2 \pi i}{N}} {\mathbbm{1}}_{n_1}, e^{\frac{4 \pi i}{N}} {\mathbbm{1}}_{n_2}, \ \dots \ , e^{\frac{2 \pi i (N-1)}{N}} {\mathbbm{1}}_{n_{N-1}})
\end{equation}
where $n_i$ are the numbers of fractional branes at that point and $n=\sum_{i=0}^{N-1} n_i$ is the total number of branes.

The spectrum of the orbifolded theory is obtained by only keeping string states that are singlets under the orbifold action.
We can build models that contain D3-branes by placing these at fixed points under the orbifold action. Such a setup is invariant under the orbifold twist and no image branes need to be introduced. The low energy spectrum on each stack of D3-branes is then given by all massless string states with endpoints on this fixed point. It is a $\mathcal{N}=1$ $\prod_{i=0}^{N-1} \textrm{U}(n_i)$ gauge theory with bifundamental matter in chiral multiplets. In detail the orbifold projection is given by
\begin{align}
\lambda=& \ {\gamma}_{\theta} \lambda {\gamma}_{\theta}^{-1} & \textrm{for gauge bosons,}\\
\lambda=& \ e^{2 \pi i {\theta}_i} {\gamma}_{\theta} \lambda {\gamma}_{\theta}^{-1} & \textrm{for complex scalars and}\\
\lambda=& \ e^{2 \pi i (\sum_i {\theta}_i s_i)} {\gamma}_{\theta} \lambda {\gamma}_{\theta}^{-1} & \textrm{for fermions,}
\end{align}
where $\lambda$ is the $n \times n$ CP matrix. The vector $\mathbf{s}$ represents the RR ground state and its entries take the values $s_i = \pm 1/2$. The GSO projection only allows states with $\sum_i s_i=\textrm{ odd}$.

In the closed string picture the introduction of orbifold twists ${\theta}^k$ leads to the presence of twisted sectors. Sectors are labelled by the amount of supersymmetry preserved, namely $\mathcal{N}=4$, $\mathcal{N}=2$ and $\mathcal{N}=1$ for the cases that three tori,
one torus or no torus are left fixed by the geometric orbifold action.


\subsection{D3-D3 string amplitudes and the background field method}
We will employ the background field method to compute the kinetic mixing from a string theory calculation. A magnetic field is turned on in the 23-direction of the non-compact space and the 1-loop vacuum amplitude is calculated in this backgrond. The magnetic field is defined as $F_{23}^a=B Q^a$ where $a$ denotes the gauge group of interest and $Q^a$ is a U(1) generator within this group.\footnote{When we introduced the background field method before we defined the magnetic field slightly differently: $F_{23}=B_1 Q^a + B_2 Q^b$. This difference however is cosmetic: We can imagine that $F_{23}=B_1 Q^a + B_2 Q^b \equiv B^{\prime} Q^{\prime}$ by subsuming the two U(1) generators into one. We arrive again at the correct result for the kinetic mixing between $\textrm{U(1)}^a$ and $\textrm{U(1)}^b$ if we then expand the string vacuum amplitude to order $\mathcal{O}({B^{\prime}}^2)$ while embedding $Q^{\prime}=Q^a$ at one end of the string and $Q^{\prime}=Q^b$ at the other.} As only open strings couple to an electromagnetic field and we consider models free of O-planes, purely the annulus worldsheet will contribute to the vacuum amplitude. For open strings starting and ending on D3-branes these annulus amplitudes have already been constructed in the given background field \cite{ABD}. We decompose the full amplitude into orbifold sectors labelled by $k$. We get different expressions depending on whether the sector is fully twisted ($\mathcal{N}=1$) or keeps one complex torus fixed ($\mathcal{N}=2$):
\begin{align}
\nonumber {\mathcal{A}}_{\mathcal{N}=1}^{(k)} = & \frac{1}{4} \int_0^{\infty} \frac{\textrm{d} t}{2t} \frac{1}{(2 {\pi}^2 t)} \sum_{\alpha, \beta = 0, 1/2}
\frac{{\eta}_{\alpha \beta}}{2} \textrm{Tr} \left( \left( {\gamma}_{{\theta}^k} \otimes {\gamma}_{{\theta}^k}^{-1} \right)
\frac{i (\beta_L + \beta_R)}{2 {\pi}^2}
\frac{\vartheta \left[\begin{array}{c} \alpha \\ \beta \end{array}\right] \left(\left. \frac{i \epsilon t}{2} \right| t \right) }{\vartheta \left[\begin{array}{c} 1/2 \\ 1/2 \end{array}\right] \left( \left. \frac{i \epsilon t}{2} \right| t \right)} \right) \times \\
& \times \prod_{i=1}^3 \left( \frac{(-2 \sin (\pi {\theta}_i^k)) \ \vartheta \left[\begin{array}{c} \alpha \\ \beta + {\theta}_i^k \end{array}\right]\left(\left.0 \right| t \right)}{\vartheta \left[\begin{array}{c} 1/2 \\ 1/2 + {\theta}_i^k \end{array}\right] \left(\left.0 \right| t \right) } \right), \\
\nonumber {\mathcal{A}}_{\mathcal{N}=2}^{(k)} = & \frac{1}{4} \int_0^{\infty} \frac{\textrm{d} t}{2t} \frac{1}{(2 {\pi}^2 t)} \sum_{\alpha, \beta = 0, 1/2}
\frac{(-1)^{2 \alpha}{\eta}_{\alpha \beta}}{2} \textrm{Tr} \left( \left( {\gamma}_{{\theta}^k} \otimes {\gamma}_{{\theta}^k}^{-1} \right)
\frac{i (\beta_L + \beta_R)}{2 {\pi}^2}
\frac{\vartheta \left[\begin{array}{c} \alpha \\ \beta \end{array}\right] \left(\left. \frac{i \epsilon t}{2} \right| t \right) }{\vartheta \left[\begin{array}{c} 1/2 \\ 1/2 \end{array}\right] \left( \left. \frac{i \epsilon t}{2} \right| t \right)} \right) \times \\
& \times \frac{\vartheta \left[\begin{array}{c} \alpha \\ \beta \end{array}\right]\left(\left.0 \right| t \right)}{{\eta}^3(t)} \prod_{i=1}^2 \left( \frac{(-2 \sin (\pi {\theta}_i^k)) \ \vartheta \left[\begin{array}{c} \alpha \\ \beta + {\theta}_i^k \end{array}\right]\left(\left.0 \right| t \right)}{\vartheta \left[\begin{array}{c} 1/2 \\ 1/2 + {\theta}_i^k \end{array}\right] \left(\left.0 \right| t \right) } \right) .
\end{align}
Charges on the left and right endpoints of the string are denoted by $q_L$ and $q_R$ and we write $\beta_L=Bq_L$ and $\beta_R = B q_R$. In addition, we define:
\begin{equation}
\epsilon = \frac{1}{\pi} (\arctan \beta_L + \arctan \beta_R) .
\end{equation}
To obtain information about kinetic mixing we are mainly interested in the $\mathcal{O}(B^2)$ terms of the above amplitudes. We expand the parts depending on the magnetic field as has been done in \cite{C}:
\begin{align}
\nonumber \textrm{Tr} & \left( \left( {\gamma}_{{\theta}^k} \otimes {\gamma}_{{\theta}^k}^{-1} \right)
\frac{i (\beta_L + \beta_R)}{2 {\pi}^2}
\frac{\vartheta \left[\begin{array}{c} \alpha \\ \beta \end{array}\right] \left(\left. \frac{i \epsilon t}{2} \right| t \right) }{\vartheta \left[\begin{array}{c} 1/2 \\ 1/2 \end{array}\right] \left( \left. \frac{i \epsilon t}{2} \right| t \right)} \right) \ \underset{\mathcal{O}(B^2)}{=} \\
& \underset{\mathcal{O}(B^2)}{=} \
- \frac{B^2 t}{16 \pi^4} \times \frac{{\vartheta}^{\prime \prime} \left[\begin{array}{c} \alpha \\ \beta \end{array}\right]\left(\left.0 \right| t \right)}{{\eta}^3(t)} \textrm{Tr} \left(q_L^2 {\gamma}_{{\theta}^k} \otimes {\gamma}_{{\theta}^k}^{-1} + 2 q_L {\gamma}_{{\theta}^k} \otimes q_R {\gamma}_{{\theta}^k}^{-1} + {\gamma}_{{\theta}^k} \otimes q_R^2 {\gamma}_{{\theta}^k}^{-1} \right).
\end{align}
This can now be simplified by applying a Riemann identity for combinations of Jacobi $\vartheta$-functions \cite{ABD, Stieberger} which can be found in the appendix. In the $\mathcal{N}=1$ sectors the amplitude reduces to
\begin{equation}
{\mathcal{A}}_{\mathcal{N}=1}^{(k)} \ \underset{\mathcal{O}(B^2)}{=} \ \frac{1}{2} {\left(\frac{B}{2 \pi^2} \right)}^2 \frac{1}{4} \textrm{Tr} \left( {(q_L + q_R)}^2 {\gamma}_{{\theta}^k} \otimes {\gamma}_{{\theta}^k}^{-1}\right) \prod_{i=1}^3 (-2 \sin (\pi \theta_i^k)) \int_0^{\infty} \frac{\textrm{d} t}{2t} \frac{1}{4 \pi}  \sum_{i=1}^3 \frac{{\vartheta}^{\prime} \left[\begin{array}{c} 1/2 \\ 1/2 - {\theta}_i^k \end{array}\right]}{ \vartheta \left[\begin{array}{c} 1/2 \\ 1/2 - {\theta}_i^k \end{array}\right]}.
\end{equation}
In the partially twisted sectors the simplification goes even further. With $\mathcal{N}=2$ supersymmetry only BPS multiplets can renormalise the gauge couplings. As the string oscillators are all non-BPS the string oscillator tower cannot contribute and the combination of Jacobi $\vartheta$-functions collapses to a single number:
\begin{equation}
{\mathcal{A}}_{\mathcal{N}=2}^{(k)} \ \underset{\mathcal{O}(B^2)}{=} \ \frac{1}{2} {\left(\frac{B}{2 \pi^2} \right)}^2 \frac{1}{4} \textrm{Tr} \left( {(q_L + q_R)}^2 {\gamma}_{{\theta}^k} \otimes {\gamma}_{{\theta}^k}^{-1}\right) \prod_{i=1}^2 (-2 \sin (\pi \theta_i^k)) \int_0^{\infty} \frac{\textrm{d} t}{2t} \frac{1}{2}  .
\end{equation}


\section{Consistent orbifold models and U(1)s}
\label{consistent}
In this section we will be examining toroidal orbifold models based on ${\mathbb{T}}^6/{\mathbb{Z}}_{N}$ which will be employed in the study of kinetic mixing later. In particular, we will be interested in designing models free of gauge anomalies based on this compactification and investigate U(1) anomalies and masses. Original work on branes at orbifold singularities can be found in \cite{DouglasMoore}.


\subsection{Tadpole cancellation}
The cancellation of RR tadpoles is crucial for the consistency of the theory. Tadpoles of $\mathcal{N}=1$ fields arise in twisted sectors of the closed string picture and have to be cancelled at the singularity at which they arise. This ensures the disappearance of cubic non-Abelian anomalies in the low-energy field theory which would otherwise render the theory pathological.

As $\mathcal{N}=1$ tadpoles have to be cancelled locally we can state a general rule for their disappearance that will not depend on the global model. Twisted tadpoles are calculated by evaluating the annulus amplitude in the closed string channel in the limit $l \rightarrow \infty$ where $l$ is the closed string cylinder length. By requiring this to vanish one finds
\begin{equation}
\label{N1tadpoles}
(\prod_{i=1}^3 2 \sin{\pi k {\theta}_i} ) \textrm{Tr} \gamma_{{\theta}^k, 3} =0 \quad \ \textrm{for all } k=1, \dots, N-1.
\end{equation}
These tadpole cancellation conditions can also be derived starting with the low energy spectrum and cancelling cubic non-Abelian anomalies.

Tadpoles can also arise in $\mathcal{N}=2$ sectors and are caused by the exchange of partially twisted RR fields. As these tadpoles can escape the singularity along untwisted directions they do not have to cancel locally, but can be balanced globally when considering the full compact model. In pure D3 models this can be achieved by placing D3-branes at various singularities in the bulk and choosing the gauge groups on them carefully. Another way of cancelling these $\mathcal{N}=2$ tadpoles employs the introduction of D7-branes, which also wrap the collapsed cycles and
lift some of the restrictions on the allowed gauge groups on the D3-branes. We will discuss the calculational details of the cancellation of $\mathcal{N}=2$ tadpoles when considering specific models later in this paper.

We note that with the inclusion only of D3-branes there still remains an uncancelled overall $\mathcal{N}=4$ tadpole. There are two ways to deal with this.
As none of our calculations require the compactness of the first two complex tori, one could allow these to remain non-compact without affecting any of our
results. Alternatively, we can note
that as the physics we are interested in is related to running gauge couplings, while $\mc{N}=4$ sectors are conformal, the $\mc{N}=4$
tadpole is not relevant to the physics that we study here. So while strictly speaking the model is incomplete, it is satisfactory for our purposes.


\subsection{U(1) anomalies and masses}
\label{NAU1}
Ultimately, we will be interested in kinetic mixing between massless U(1)s as anomalous U(1)s aquire a mass at the string scale and are hence removed from low energy dynamics. Abelian gauge bosons become massive if the effective four dimensional Lagrangian contains a Green-Schwarz coupling of the form $C_2 \wedge F_i$. In our case $C_2$ comes from an RR twisted 2-form and $F$ is the field strength of the Abelian group ${\textrm{U}(1)}_i$. This coupling arises automatically for anomalous U(1)s in the course of anomaly cancellation via a Green-Schwarz mechanism and generates a string scale mass. However, a Green-Schwarz coupling to a partially twisted RR 2-form can also be generated for non-anomalous U(1)s which, in a global completion of the model, generates a mass at the KK scale. In this section we will review anomalies of U(1)s in orbifold models and identify all non-anomalous U(1)s.

The gauge theory realized on a stack of $n$ D3-branes located at a ${\mathbb{Z}}_N$ orbifold singularity contains up to $N$
U(1) group factors and up to $N$ non-Abelian group factors. In the previous section we ensured that no non-Abelian cubic anomalies remain by cancelling all $\mathcal{N}=1$ twisted tadpoles. Further anomalies, which manifest themselves as triangle diagrams in four dimensional theory, are mixed ${\textrm{U}(1)}_j\times{\textrm{G}}_l^2 $ anomalies ${\mathcal{A}}_{jl}$ where ${\textrm{G}}_l$ is a non-Abelian gauge group, as well
as cubic Abelian anomalies. These are cancelled in string theory via the Green-Schwarz coupling $C_2 \wedge F_i$ to a fully twisted RR 2-form, for which here we simply state results.

We follow the treatment of anomalies as presented in \cite{AIQU, IRU}. When considering a general U(1) at the singularity
\begin{equation}
\label{Q}
Q_c = \sum_{j=0}^{N-1} c_j \frac{Q_j}{n_j},
\end{equation}
the condition for it to be non-anomalous becomes
\begin{equation}
\sum_{j=0}^{N-1} \frac{c_j}{n_j} {\mathcal{A}}_{jl} =0 \quad \ \forall \ l .
\end{equation}
We can diagonalize this by inserting
\begin{equation}
\label{cs}
c_j = \frac{1}{N} \sum_{k=0}^{N-1} e^{- 2 \pi i j k } r_k .
\end{equation}
Using the explicit form of the mixed anomalies the condition for non-anomalous U(1)s then reads:
\begin{equation}
\label{NAU1s}
\left(\prod_{i=1}^3 2 \sin{\pi k {\theta}_i}  \right) r_k =0,
\end{equation}
where ${\theta}_i$ again is the orbifold twist. There is always one trivial solution where $r_0$ is arbitrary and all other $r_k=0$, but there are further solutions if the prefactor vanishes for any additional $k$ apart from zero. This is indeed the case if we have $\mathcal{N}=2$ sectors in the orbifold projection and we get one further non-anomalous U(1) for each $\mathcal{N}=2$ sector. The trivial solution is referred to as the diagonal U(1) in the literature and is defined as
\begin{equation}
Q_{diag}=\sum_{i=0}^{N-1} \frac{Q_i}{n_i} .
\end{equation}
We will be more interested in the additional non-anomalous U(1)s which are given by further solutions to equation \ref{NAU1s}. Without loss of generality the parameters $r_k$ can be chosen such that the additional non-anomalous U(1)s are orthogonal to $Q_{diag}$.

Now that we have identified all non-anomalous U(1)s in our orbifold model we need to examine whether they are massless. As this will depend on the global properties of the model, we will perform this examination in a specific model.


\section{Kinetic mixing: a $\mathbb{T}^6 / {\mathbb{Z}}_{4}$ example}
\label{kinmixD3}

We will now explore the possibilities for kinetic mixing in a simple model based on a ${\mathbb{T}}^6/{\mathbb{Z}}_{4}$ orbifold with D3 branes located at orbifold fixed points. This model will allow us to identify the
necessary conditions for kinetic mixing which can be applied to more intricate models later.


\subsection{Fully and partially twisted tadpoles}

\FIGURE[ht]{
\label{Z4}
\includegraphics[width=0.75\textwidth]{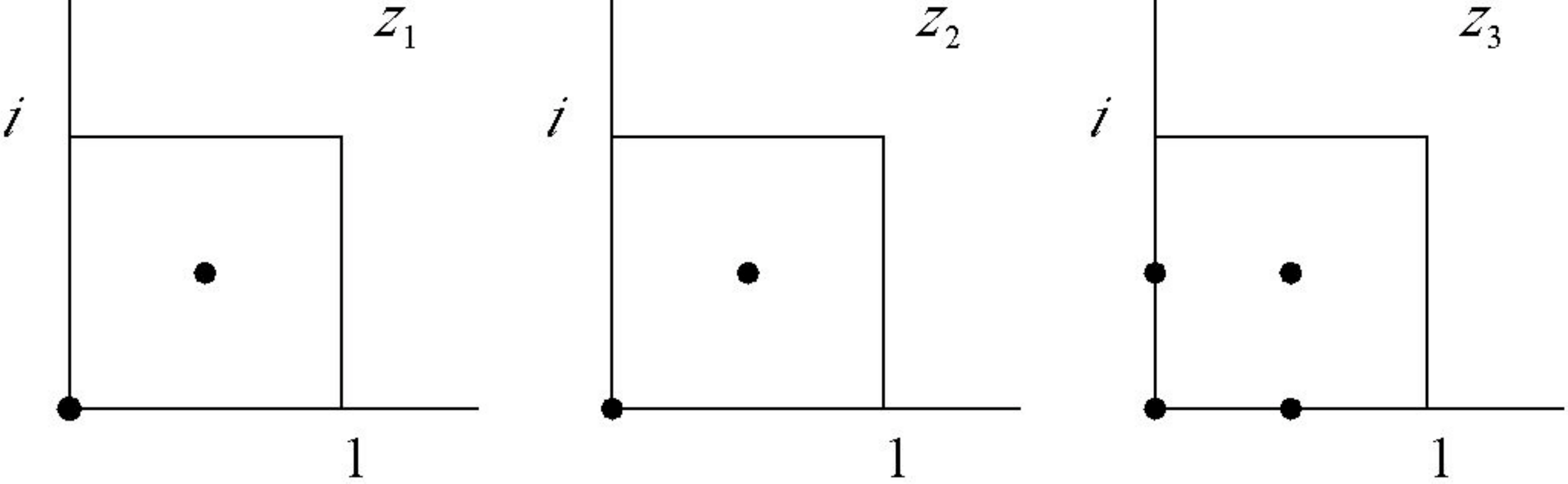}
\caption{The $\mathbb{T}^6/\mathbb{Z}_4$ orbifold. The circles indicate the positions of the 16 orbifold fixed points.}
}
The ${\mathbb{T}}^6/{\mathbb{Z}}_{4}$ orbifold is generated by the action $\theta = \frac{1}{4}(1,1,-2)$ where each twist acts on one complex plane of the compact space. The resulting orbifold group is
\begin{equation}
\left\{ \left(0,0,0\right), \left(\frac{1}{4},\frac{1}{4},-\frac{1}{2}\right), \left(\frac{1}{2},\frac{1}{2},-1\right), \left(\frac{3}{4},\frac{3}{4},-\frac{3}{2}\right) \right\} .
\end{equation}
The compact space is factorized into three two-tori which are defined by identification on the SU(2) root lattice to be consistent with crystallographic restriction.\footnote{The orbifold element acting on the third two-torus is an element of $\mathbb{Z}_2$ in this case. Crystallographic restriction does not require this torus to be defined on a SU(2) root lattice. In fact, a torus with any complex structure modulus $U$ is allowed.}

There are 16 fixed points of the orbifold action on the compact space which can be seen in figure \ref{Z4}. Each of these fixed points is an orbifold singularity and in orbifold models we can place D3-branes at singularities without the need for image branes.

We are now in a position to build a model. To do so we position stacks of D3-branes at the following four orbifold fixed points that only differ in the third complex coordinate:
\begin{equation}
A = \Big(0,0,0\Big),  \ B = \Big(0,0,\frac{i}{2}\Big), \ C = \Big(0,0,\frac{1}{2}\Big), \ D = \Big(0,0,\frac{1+i}{2}\Big).
\end{equation}
In addition, we pick an embedding of the orbifold action on the Chan-Paton factors for each stack separately: $${\gamma}_{\theta}^I = \left( {\mathbbm{1}}_{n_0^I}, i {\mathbbm{1}}_{n_1^I}, - {\mathbbm{1}}_{n_2^I}, -i {\mathbbm{1}}_{n_3^I} \right),$$ where $I$ is a label for either fixed point. To render the model consistent fully twisted tadpoles have to be cancelled at each fixed point individually. The $\mathbb{Z}_4$ orbifold has two $\mathcal{N}=1$ sectors generated by ${\theta}^1$ and ${\theta}^3$ which are identical with regard to the conditions they set. As there are neither D7 nor O-planes branes present the $\mathcal{N}=1$ tadpole cancellation conditions translate as $$\textrm{Tr} {\gamma}_{\theta}^{I} =0,$$ leading to the conditions $n_0^I=n_2^I$ and $n_1^I=n_3^I$ at all four fixed points.

\FIGURE[ht]{
\label{stringdiags}
\includegraphics[width=0.75\textwidth]{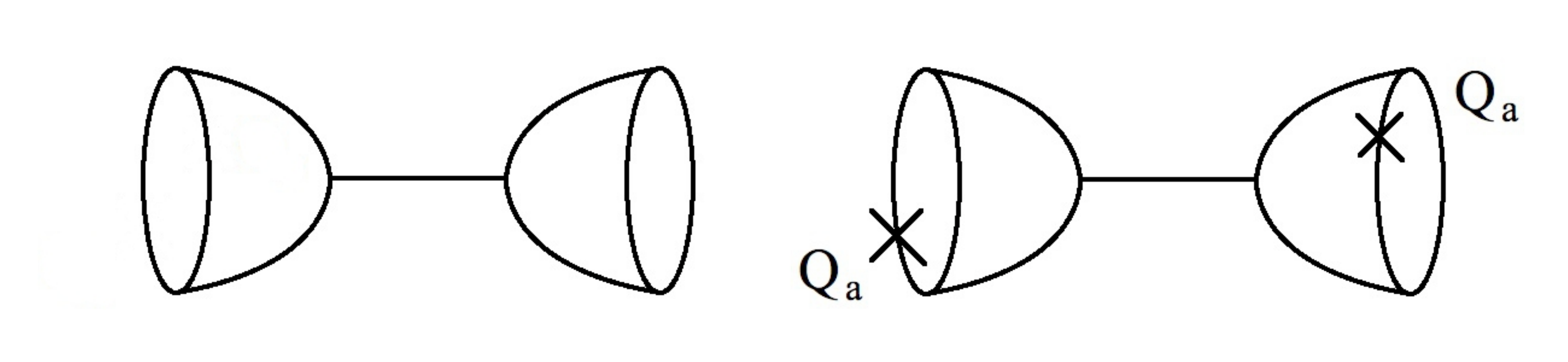}
\caption{The annulus string diagram in the open string UV/ closed string IR limit $l =1/t \rightarrow \infty$. Both ends of the left diagram are proportional to tadpoles. The right diagram has insertions of a U(1) generator on both its boundaries and is used to calculate Green-Schwarz couplings which give a mass to U(1) gauge bosons.}}
As we are working in a global model we also have to ensure the vanishing of partially twisted tadpoles in the $\mathcal{N}=2$ sector. In the $\mathbb{Z}_4$ orbifold, partially twisted tadpoles arise in the $\theta^2$ sector which leaves the third 2-torus invariant. In homology, the corresponding 2-cycle therefore links all four fixed points which differ by their location in $z_3$ and we can achieve tadpole cancellation by balancing them across the fixed points.

We can calculate tadpoles by considering a string diagram as shown on the left in figure \ref{stringdiags} which is a vacuum annulus diagram evaluated in the limit where the cylinder length $l=1/t$ gets large. It corresponds to a vacuum tadpole sourcing a partially twisted massless RR mode which propagates before it is absorbed again by a vacuum tadpole. The partially twisted RR mode arises in the $\theta^2$ sector and travels across the third 2-torus, linking tadpoles arising at all four fixed points A, B, C and D. To account for this in the string calculation we need to evaluate the annulus diagram for all 16 combinations of strings starting and ending on either of the four fixed points A, B, C and D. As we are analysing the situation in a compact model we also need to consider strings which wrap the compact space multiple times while stretching between the various fixed points. However, in the limit $l \rightarrow \infty$ the sum over winding modes collapses to a factor common to all string states wrapping the third two-torus. Hence, the annulus amplitude can be writen as
\begin{equation}
{\mathcal{A}}_{\mathcal{N}=2} \ \underset{l \rightarrow \infty}{\propto} \ \bigg(\sum_{\substack{I= \\A,B,C,D }} \textrm{Tr} \left({\gamma}_{{\theta}^2}^I \right)\bigg) \bigg(\sum_{\substack{J= \\A,B,C,D }}  \textrm{Tr} \left({{\gamma}_{{\theta}^2}^J}^{*}\right)\bigg) \prod_{i=1}^2 (-2 \sin (\pi \theta_i^2)) \int_{l^{\prime}}^{\infty} \textrm{d} l,
\end{equation}
where the sums over $I$ and $J$ are over the orbifold fixed points with D-branes placed on them. The integral produces a linear divergence and we need to cancel it in a consistent model. We turn our attention to the sums over the Chan-Paton traces which factorise into tadpole contributions coming from the left-hand and from the right-hand sides of the diagram. The vanishing of the above amplitude can be achieved by requiring
\begin{equation}
\sum_{\substack{I= \\A,B,C,D }} \textrm{Tr} \left({\gamma}_{{\theta}^2}^I \right) = \textrm{Tr} {\gamma}_{{\theta}^2}^{A} + \textrm{Tr} {\gamma}_{{\theta}^2}^{B} + \textrm{Tr} {\gamma}_{{\theta}^2}^{C} + \textrm{Tr} {\gamma}_{{\theta}^2}^{D} =0,
\end{equation}
which cancels partially twisted tadpoles coming from either end of the diagram individually.

Having identified the conditions for both fully and partially twisted tadpoles to cancel we are now in a position to construct a consistent model with D3-branes at $\mathbb{Z}_4$ orbifold singularities. Once the model is presented we will identify the massless U(1)s present.

We cancel tadpoles by embedding the orbifold action on the Chan-Paton factors as follows: Written as vectors $\vec{n}$ the numbers of fractional branes are
\begin{eqnarray}
{\vec{n}}^A &=& (N,M,N,M) \\
{\vec{n}}^B &=& (M,N,M,N) \\
{\vec{n}}^C &=& (K,L,K,L) \\
{\vec{n}}^D &=& (L,K,L,K)
\end{eqnarray}
where $N$, $M$, $K$ and $L$ are positive integers. Fully twisted tadpoles vanish since the above ensures $n_0=n_2$ and $n_1=n_3$. Partially twisted tadpoles cancel as
$$ \textrm{Tr} {\gamma}_{{\theta}^2}^{A} + \textrm{Tr} {\gamma}_{{\theta}^2}^{B} + \textrm{Tr} {\gamma}_{{\theta}^2}^{C} + \textrm{Tr} {\gamma}_{{\theta}^2}^{D} = 2(N-M) - 2(N-M) + 2(K-L) - 2(K-L) =0 .$$


\subsection{Massless and massive non-anomalous U(1)s}

After having assigned fractional branes and cancelled twisted tadpoles we can now study the $U(1)$s that are present. In particular, each stack of branes supports a consistent gauge theory with gauge group $\textrm{U}(n_0) \times \textrm{U}(n_1) \times \textrm{U}(n_2) \times \textrm{U}(n_3)$. Locally, we can identify $\textrm{U}(N)=\textrm{SU}(N) \times \textrm{U}(1)$ and our model thus automatically contains a multitude of U(1) factors which are central to the present work. However, not all of these U(1) factors will be of interest to us since most of them will be anomalous and hence acquire a mass by the Green-Schwarz mechanism. Only certain combinations of the various U(1) factors on a stack of D3-branes will remain non-anomalous and it is these combinations we will study. We recall that for each orbifold singularity there will be at least one non-anomalous U(1) on each stack of D3-branes which is termed the diagonal combination. In addition, for each $\mathcal{N}=2$ sector corresponding to a different two-cycle there exists one further non-anomalous combination of U(1)s on each stack of D3-branes. The calculational details of this analysis are described in section \ref{NAU1}. In the context of the ${\mathbb{Z}}_4$ orbifold the above is realised as follows. As the ${\mathbb{Z}}_4$ orbifold displays one $\mathcal{N}=2$ sector we can identify two non-anomalous U(1)s on the worldvolume of each stack of D3-branes:
\begin{eqnarray}
{\textrm{U}(1)}_{diag} &=& \frac{1}{n_0}({\textrm{U}(1)}_0 + {\textrm{U}(1)}_2)+ \frac{1}{n_1}({\textrm{U}(1)}_1 + {\textrm{U}(1)}_3),  \\
{\textrm{U}(1)}_{tw} &=& \frac{1}{n_1}({\textrm{U}(1)}_0 + {\textrm{U}(1)}_2)- \frac{1}{n_0}({\textrm{U}(1)}_1 + {\textrm{U}(1)}_3),
\end{eqnarray}
where the two Abelian generators are chosen to be orthogonal.

Before proceeding with the kinetic mixing calculation we have to examine whether any of the non-anomalous U(1)s become massive. A study of non-anomalous U(1)s in the context of the $\mathbb{Z}_4$ orbifold has been presented in \cite{CP} and we will follow this analysis.
 Despite being free of anomalies, the Abelian gauge bosons can still gain a mass due to a non-vanishing Green-Schwarz coupling to a partially twisted RR mode. In particular, for the U(1) to remain massless the string diagram shown on the right-hand side of figure \ref{stringdiags} has to vanish. It corresponds to the annulus vacuum amplitude expanded to second order in the background field $B$ and evaluated for large cylinder length $l=1/t \rightarrow \infty$. We consider the diagram with one U(1) generator inserted on the left boundary of the annulus and another on the right boundary. Both ends of the diagram represent Green-Schwarz couplings between the U(1) and a massless RR mode which propagates between them. Specifically, we embed the charges of the string endpoints within the U(1)-generator in question:
\begin{align}
q_L=-q_R= Q_{diag}=&\frac{1}{{\mathcal{N}}_{diag}} \left(\frac{\mathbbm{1}_{n_0}}{n_0} , \frac{\mathbbm{1}_{n_1}}{n_1} , \frac{\mathbbm{1}_{n_0}}{n_0} , \frac{\mathbbm{1}_{n_1}}{n_1} \right) & \textrm{or}\\
q_L=-q_R=Q_{tw}= &\frac{1}{{\mathcal{N}}_{diag}} \left(\frac{\mathbbm{1}_{n_0}}{n_1} ,-\frac{\mathbbm{1}_{n_1}}{n_0} ,\frac{\mathbbm{1}_{n_0}}{n_1} ,-\frac{\mathbbm{1}_{n_1}}{n_0} \right)&
\end{align}
where we were careful to include the normalisations such that $\textrm{Tr} \ Q^2=1$.
The statement that a U(1) is non-anomalous is equivalent to the fact that it has a vanishing Green-Schwarz coupling in the $\mathcal{N}=1$ sectors.\footnote{In equations (4.9) and (4.10) we identified non-anomalous U(1)s in the spectrum on the D3-branes based on 4D field theory techniques. Alternatively one can perform the same analysis in string theory: U(1)s are then free of anomalies if their Green-Schwarz couplings vanish locally in all $\mathcal{N}=1$ sectors.} Hence we only need to evaluate the above diagram in the $\mathcal{N}=2$ sector to check whether any of the anomaly-free U(1)s acquire a mass. In the $\mathcal{N}=2$ sector the RR-mode is allowed to propagate across the third two-torus and hence we will have to include winding states in our expression. As the U(1)s are defined on a single stack of branes so far, we only have to consider strings that start and end on the same orbifold singularity. In the limit $l=1/t \rightarrow \infty$ the sum over winding modes collapses to a single factor:
\begin{equation}
\sum_{n,m}= e^{- \pi R^2 (n^2 + m^2)t} \rightarrow \frac{l}{R^2}\left(1+ \mathcal{O}(e^{-\frac{\pi}{R^2}l}) \right) .
\end{equation}
Setting $q_L=-q_R=Q$ the amplitude therefore becomes
\begin{equation}
{\mathcal{A}}_{\mathcal{N}=2}^{q_L q_R} \ \underset{l \rightarrow \infty}{\propto} \ {\left| \textrm{Tr} \left(Q \ {\gamma}_{{\theta}^2} \right) \right|}^2 \ \prod_{i=1}^2 (-2 \sin (\pi \theta_i^2)) \int_{l^{\prime}}^{\infty} \frac{\textrm{d} l}{l} \frac{l}{R^2} .
\end{equation}
Hence, for a non-anomalous U(1) to remain massless in the context of the $\mathbb{Z}_4$ orbifold we find that the trace factor has to vanish in the $\mathcal{N}=2$ sector:
\begin{equation}
\textrm{Tr} \left(Q \ {\gamma}_{{\theta}^2} \right)=0.
\end{equation}
It can be easily verified that ${\textrm{U}(1)}_{diag}$ is trivially massless as $\textrm{Tr} \left( Q_{diag}{\gamma}_{{\theta}^k} \right) =0$ for all sectors labelled by $k=1,2,3$.
In the case of ${\textrm{U}(1)}_{tw}$ the expression $\textrm{Tr} \left( Q_{tw}{\gamma}_{{\theta}^k} \right)$ is only zero for the $\mathcal{N}=1$ sectors $k=1,3$ which is the statement that ${\textrm{U}(1)}_{tw}$ is non-anomalous. However in the $\mathcal{N}=2$ sector the trace is non-zero and ${\textrm{U}(1)}_{tw}$ gains a mass at the scale $M_s/R$. This analysis is valid for each orbifold singularity individually. We conclude that on each stack of D3-branes we can identify two non-anomalous U(1)s, one of which is massless while the other gains a mass at the KK scale.
However we will shortly see that combinations of $U(1)$s across singularities may remain massless.

We have now arrived at a consistent string model based on a toroidal orbifold ${\mathbb{T}}^6/{\mathbb{Z}}_{4}$ and have identified the non-anomalous U(1)s that are realized at each singularity. We are now
in a position to study whether kinetic mixing occurs in our setup.


\subsection{Prerequisites for kinetic mixing}

\FIGURE[ht]{
\label{mixdiag}
\includegraphics[width=0.75\textwidth]{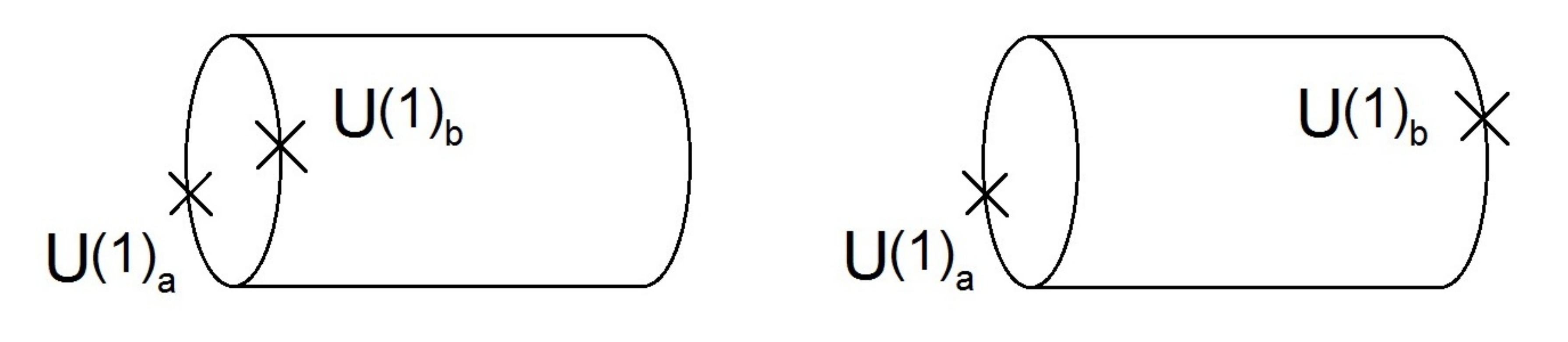}
\caption{String diagrams that contribute to mixing between two U(1)s labelled by $a$ and $b$ at $\mathcal{O}(B^2)$.}}
A mixing term between two different Abelian gauge bosons originates from string diagrams shown in figure \ref{mixdiag} at 1-loop in string theory. In the low-energy field theory this mixing can either emerge as kinetic mixing or mass mixing. To distinguish kinetic mixing from mass mixing we recall that the latter is sourced purely by massless strings whereas massive strings can contribute to the former \cite{Durham}.

Phenomenologically, kinetic mixing between hidden U(1)s is the most interesting. Two U(1)s are hidden from one another
if there is no open string state in the low energy theory that is charged under both Abelian groups simultaneously. This can be achieved easily by realizing the U(1)s on separate singularities in the compact space. Strings charged under both groups necessarily have to stretch from one singularity to the other in the compact space and hence become massive thus disappearing from the low energy spectrum.

The only massless U(1) group realized at a single singularity is the diagonal combination with generator $Q_{diag}$. When examining the mixing terms in the string calculation there are the following group theoretical factors which always vanish for the diagonal U(1)s
\begin{equation}
\textrm{Tr} \left[ Q_{diag}^I {\gamma}_{{\theta}^k}^I \right] =0,
\end{equation}
where $I$ stands for any orbifold fixed point. The trace factor is the same that appeared when we calculated the Green-Schwarz couplings; hence it is no accident that the diagonal U(1) does not exhibit kinetic mixing. Due to the emergence of the same trace factors we conclude that massless U(1)s realized at single orbifold fixed points do not mix in our $\mathbb{Z}_4$ example. We will be able to generalise this statement in the course of this work.

Instead, we will now shift our attention to Abelian groups that stretch over at least two different singularities. We can safely ignore any combinations that involve the diagonal generator as any linear combination of these will still lead to vanishing mixing. Hence we are left with linear combinations of ${\textrm{U}(1)}_{tw}$. Although ${\textrm{U}(1)}_{tw}$ restricted to a single singularity becomes massive, we can find linear combinations across several singulaities that are actually massless. As a matter of prudence we allocated D3 branes to the various singularities to allow three mutually orthogonal massless U(1)s:
\begin{eqnarray}
\nonumber {\textrm{U}(1)}_X &=& \frac{1}{\mathcal{N}_X} \left({\textrm{U}(1)}_{tw}^A - {\textrm{U}(1)}_{tw}^B \right)\\
\label{U1s} {\textrm{U}(1)}_Y &=& \frac{1}{\mathcal{N}_Y} \left({\rm{U}(1)}_{tw}^C - {\textrm{U}(1)}_{tw}^D \right)\\
\nonumber{\textrm{U}(1)}_Z &=& \frac{1}{\mathcal{N}_Z} \left[\frac{1}{\alpha} \left({\textrm{U}(1)}_{tw}^A + {\textrm{U}(1)}_{tw}^B \right) + \frac{1}{\beta} \left({\textrm{U}(1)}_{tw}^C + {\textrm{U}(1)}_{tw}^D \right)\right]
\end{eqnarray}
where $\mathcal{N}_X$,$\mathcal{N}_Y$ and $\mathcal{N}_Z$ are normalizations such that $\textrm{Tr} \ Q^2=1$ and $\alpha=(\frac{N}{M}+\frac{M}{N})$ and $\beta=-(\frac{K}{L}+\frac{L}{K})$ such that ${\textrm{U}(1)}_Z$ is massless.

There is an advantage in choosing these combinations as the basis for our examination of kinetic mixing: not only are these three Abelian generators mutually orthogonal, but ${\textrm{U}(1)}_X$ and ${\textrm{U}(1)}_Y$ are also obviously hidden from one another as only massive open strings are charged under both groups. In principle uncharged massless closed strings could be exchanged between the two brane stacks via a Green-Schwarz coupling, but by ensuring the $U(1)$s are massless we also ensured there is no such coupling present.
This is exactly the situation we wanted to explore and we now proceed to calculate mixing between ${\textrm{U}(1)}_X$ and ${\textrm{U}(1)}_Y$ using the background field method.


\subsection{Kinetic mixing between hidden U(1)s}

We assign the string endpoint charges as $q_L=Q_X$ and $q_R=Q_Y$. Our interest is in the term in the expansion of the string vacuum amplitude ${\mathcal{A}}_{\mathcal{N}=2}^{q_L q_R}$ which corresponds to the diagram shown on the right in figure \ref{mixdiag}. From the definition of ${\textrm{U}(1)}_X$ and ${\textrm{U}(1)}_Y$ it is obvious that strings that are charged under both U(1)s have to link the following pairs of fixed points: AC, AD, BC and BD. The trace factors that arise in the amplitude vanish for all $\mathcal{N}=1$ sectors as both the U(1)s are non-anomalous. However, in the $\mathcal{N}=2$ sectors we get the following non-zero results:
\begin{align}
AC: & \qquad \textrm{Tr} \left[ Q_{tw}^A {\gamma}_{{\theta}^2}^{A} \right] \cdot \textrm{Tr} \left[ Q_{tw}^C{{\gamma}^{*}}_{{\theta}^2}^{C} \right] &=& \ \frac{1}{\mathcal{N}_X \mathcal{N}_Y} \left(\frac{N}{M} + \frac{M}{N} \right) \cdot \left(\frac{K}{L} + \frac{L}{K} \right) \\
AD: & \qquad \textrm{Tr} \left[ Q_{tw}^A {\gamma}_{{\theta}^2}^{A} \right] \cdot \textrm{Tr} \left[ -Q_{tw}^D{{\gamma}^{*}}_{{\theta}^2}^{D} \right] &=& \ \frac{-1}{\mathcal{N}_X \mathcal{N}_Y} \left(\frac{N}{M} + \frac{M}{N} \right) \cdot \left(\frac{K}{L} + \frac{L}{K} \right) \\
BC: & \qquad \textrm{Tr} \left[ -Q_{tw}^B {\gamma}_{{\theta}^2}^{B} \right] \cdot \textrm{Tr} \left[ Q_{tw}^C{{\gamma}^{*}}_{{\theta}^2}^{C} \right] &=& \ \frac{-1}{\mathcal{N}_X \mathcal{N}_Y} \left(\frac{N}{M} + \frac{M}{N} \right) \cdot \left(\frac{K}{L} + \frac{L}{K} \right) \\
BD: & \qquad \textrm{Tr} \left[ -Q_{tw}^B {\gamma}_{{\theta}^2}^{B} \right] \cdot \textrm{Tr} \left[ -Q_{tw}^D{{\gamma}^{*}}_{{\theta}^2}^{D} \right] &=& \ \frac{1}{\mathcal{N}_X \mathcal{N}_Y} \left(\frac{N}{M} + \frac{M}{N} \right) \cdot \left(\frac{K}{L} + \frac{L}{K} \right).
\end{align}

The remaining part of the string vacuum amplitude, the oscillator sum, collapses to a constant factor for all $\mathcal{N}=2$ sectors
as string oscillators are non-BPS.

Since we are working in a global model, strings that stretch from one singularity to another can actually wrap the compact space a multiple times and we also have to include these winding states in our analysis. Stretching a string changes its mass and hence the winding states enter the expression as a modification of the mass in the CFT calculation. Thus strings connecting two fixed points in the compact space come with the following factors accounting for the winding modes:
\begin{eqnarray}
AC: & \qquad & \sum_{n,m = -\infty}^{\infty} e^{- \pi \left( n^2 + {\left( m + \frac{1}{2} \right)}^2 \right) R^2 t}, \\
AD: & \qquad & \sum_{n,m = -\infty}^{\infty} e^{- \pi \left( {\left( n + \frac{1}{2} \right)}^2 + {\left( m + \frac{1}{2} \right)}^2 \right) R^2 t}, \\
BC: & \qquad & \sum_{n,m = -\infty}^{\infty} e^{- \pi \left( {\left( n + \frac{1}{2} \right)}^2 + {\left( m + \frac{1}{2} \right)}^2 \right) R^2 t}, \\
BD: & \qquad & \sum_{n,m = -\infty}^{\infty} e^{- \pi \left( n^2 + {\left( m + \frac{1}{2} \right)}^2 \right) R^2 t}.
\end{eqnarray}

Putting these results together we arrive at the following expression for the kinetic mixing part of the string vacuum amplitude:
\begin{align}
\nonumber {\mathcal{A}}_{\mathcal{N}=2}^{q_L q_R} \quad \underset{\mathcal{O}(B^2)}{=}& \quad \frac{1}{4} \frac{1}{\mathcal{N}_X \mathcal{N}_Y} \left(\frac{N}{M} + \frac{M}{N} \right) \cdot \left(\frac{K}{L} + \frac{L}{K} \right) {\left( \frac{B}{2 {\pi}^2} \right)}^{2} \times \\ &
\times \int_0^{\infty} \frac{\textrm{d} t}{2t} \sum_{n,m = -\infty}^{\infty} \left(e^{- \pi \left( n^2 + {\left( m + \frac{1}{2} \right)}^2 \right) R^2 t} - e^{- \pi \left( {\left( n + \frac{1}{2} \right)}^2 + {\left( m + \frac{1}{2} \right)}^2 \right) R^2 t} \right).
\end{align}

The above expression consists of numerical prefactors determined by the number of branes and hence gauge groups present and a possible divergent part in form of an integral of the winding modes over the modular parameter $t$. For this to be valid and kinetic mixing to occur we need the above expression to be finite. Consequently, we will examine the integrand for divergences.

It is easy to see that in the open string IR regime when $t$ is large the amplitude vanishes as follows $${\mathcal{A}}_{\mathcal{N}=2}^{q_L q_R} \underset{t^{\prime} \rightarrow \infty}{\propto} \int_{t^{\prime}}^{\infty} \frac{\textrm{d}t}{t} \ \mathcal{O}\left(e^{-\frac{\pi R^2 t}{4}}\right) \rightarrow 0 .$$
This is to be expected as by construction there are no states in the open string low energy spectrum that are charged under the two U(1)s since the Abelian groups are hidden from one another.

To investigate the behaviour in the open string UV regime of small $t$ the integrand is rewritten using Poisson resummation giving an equivalent expression for the kinetic mixing amplitude:
\begin{align}
\nonumber {\mathcal{A}}_{\mathcal{N}=2}^{q_L q_R} \quad \underset{\mathcal{O}(B^2)}{=}& \quad \frac{1}{4} \frac{1}{\mathcal{N}_X \mathcal{N}_Y} \left(\frac{N}{M} + \frac{M}{N} \right) \cdot \left(\frac{K}{L} + \frac{L}{K} \right) {\left( \frac{B}{2 {\pi}^2} \right)}^{2} \times \\ &
\times \int_0^{\infty} \frac{\textrm{d} t}{2t} \frac{1}{R^2 t} \sum_{n,m = -\infty}^{\infty} {\left(-1\right)}^m \ \left[1 - {\left(-1\right)}^n \right] \ e^{- \left( n^2 +m^2 \right) \frac{\pi}{R^2 t}} .
\end{align}
Hence we find that the kinetic mixing amplitude also vanishes in the UV limit:
$${\mathcal{A}}_{\mathcal{N}=2}^{q_L q_R} \underset{t^{\prime} \rightarrow 0}{\propto} \int_{0}^{t^{\prime}} \frac{\textrm{d}t}{t} \ \mathcal{O}\left(e^{-\frac{\pi}{ R^2 t}}\right) \rightarrow 0 .$$

In fact, the integrand does not diverge anywhere and can be integrated to give a finite result
which is precisely the kinetic mixing. Hence we established the presence of kinetic mixing between ${\textrm{U}(1)}_X$ and ${\textrm{U}(1)}_Y$.

Having identified this effect our analysis of kinetic mixing in the context of the $\mathbb{Z}_4$ is not yet complete. In our ${\mathbb{Z}}_4$ orbifold model we identified three massless and mutually orthogonal U(1)s which could engage in kinetic mixing \eqref{U1s}. At most, there could be mixing between all three U(1)s. We just considered the case when the two mutually hidden U(1)s mixed. In the other possible instances the U(1)s involved are not necessarily hidden from one another as there exist massless strings that are charged under both of them. As it turns out, the only mixing in our model is between ${\textrm{U}(1)}_X$ and ${\textrm{U}(1)}_Y$ and any mixing between the others vanishes. This is not a general observation but a feature of our particular model and the symmetries present in the distribution of fractional branes across fixed points.


\subsection{Size of kinetic mixing}
Having established the appearance of kinetic mixing between two massless U(1)s in a toroidal orbifold model based on ${\mathbb{T}}^6/{\mathbb{Z}}_{4}$ we now want to assess the strength of this effect. We can extract the kinetic mixing parameter ${\chi}$ from the vacuum string amplitude ${\mathcal{A}}_{\mathcal{N}=2}^{q_L q_R}$:
\begin{equation}
{\left. \frac{{\chi}}{g_X g_Y} \right|}_{\textrm{1-loop}} = \frac{1}{4 {\pi}^2} {\Lambda}_2,
\end{equation}
where
\begin{equation}
{\mathcal{A}}_{\mathcal{N}=2}^{q_L q_R} \ \underset{\mathcal{O}(B^2)}{=} \ \frac{1}{2} {\left( \frac{B}{2 {\pi}^2} \right)}^{2} \ {\Lambda}_2 .
\end{equation}
To arrive at a numerical value we will perform the integral in the expression for ${\Lambda}_2$.  As the result does not diverge we are allowed to swap the order of the integration and the sum over winding modes. We will make use of the fact that we have two equivalent expressions for the kinetic mixing parameter and integrate piecewise over the Poisson resummed expression for small $t<1/R^2$ and the original one for large $t> 1/R^2$:
\begin{align}
\nonumber \chi \propto &
\sum_{n,m = -\infty}^{\infty} \int_0^{\frac{1}{R^2}} \frac{\textrm{d} t}{t} \ \frac{1}{R^2 t} \ {\left(-1\right)}^m \ \left[1 - {\left(-1\right)}^n \right] \ e^{- \left( n^2 +m^2 \right) \frac{\pi}{R^2 t}} \ +\\
& + \sum_{n,m = -\infty}^{\infty} \int_{\frac{1}{R^2}}^{\infty} \frac{\textrm{d} t}{t}  \left(e^{- \pi \left( n^2 + {\left( m + \frac{1}{2} \right)}^2 \right) R^2 t} - e^{- \pi \left( {\left( n + \frac{1}{2} \right)}^2 + {\left( m + \frac{1}{2} \right)}^2 \right) R^2 t} \right)
\end{align}
where we suppressed the numerical prefactors. To give a numerical estimate of the kinetic mixing parameter we then only need to consider the lowest winding modes to get a good approximation. The integration leaves us with the following result that includes the exponential integral function $\textrm{E}_1$:
\begin{align}
\nonumber \chi \propto &
\sum_{n,m = -\infty}^{\infty} \left\{ \frac{{\left(-1\right)}^m \ \left[1 - {\left(-1\right)}^n \right]}{\pi \left( n^2 +m^2 \right)} \  e^{- \pi \left( n^2 + m^2 \right)} \right. + \\
+ & \left. {\textrm{E}}_1 \left(\pi \left[ n^2 + { (m + \frac{1}{2} )}^2 \right] \right) - {\textrm{E}}_1 \left(\pi \left[{ (n + \frac{1}{2} )}^2 + { (m + \frac{1}{2} )}^2 \right] \right) \right\}
\end{align}

As the strength of kinetic mixing is now evidently independent of $R^2$ we are in a position to evaluate the above for a compact space of general volume. The sum over winding modes can be evaluated numerically and contributes a factor of the order $\mathcal{O}(1)$. The numerical prefactor of the kinetic mixing parameter depends on the model, specifically, on the numbers of D-branes and hence the gauge groups present. We can evaluate this factor for phenomenologically interesting gauge groups and arrive at an estimate for the strength of kinetic mixing:
\begin{equation}
10^{-3} \lesssim \left\{ \frac{{\chi}}{g_X g_Y} \right\} \lesssim 10^{-1} .
\end{equation}
The size of the kinetic mixing effect is consistent with previous expectations \cite{JMR}.


\subsection{Dependence on the moduli}
In the numerical analysis of the kinetic mixing in our ${\mathbb{Z}}_4$ orbifold model we observed the disappearance of $R^2$ from our final expression. This quantity is proportional to the volume of the compact space which is identical to the imaginary part of the K\"{a}hler modulus of that two-torus. We arrive at the interesting result that in this specific example kinetic mixing is ignorant of the volume of the compact space being wrapped. This is an observation which we would like to generalise. Consequently, we will study the sensitivity of the kinetic mixing parameter to the complex and K\"{a}hler moduli more systematically. The dependence on the moduli enters via the sum over winding modes. The winding factor for strings wrapping a two-torus can be written as
\begin{equation}
\sum_{n,m} e^{- \pi \frac{T_2}{U_2} {\left|z + U n + m \right|}^2 t},
\end{equation}
where $z$ is the complex separation between the endpoints of the open string, $U$ is the complex modulus, $T$ the K\"{a}hler modulus and the quantities with subscript 2 are their respective imaginary parts. We can again obtain an equivalent Poisson resummed expression given by
\begin{equation}
\sum_{q,p} \frac{1}{T_2 t} e^{- \frac{\pi}{T_2 U_2 t} {\left|q + \bar{U} p \right|}^2 + \frac{2 \pi i}{U_2} \textrm{Im} \left( z (q + \bar{U} p) \right) } .
\end{equation}
We can perform the integral over the modular parameter $t$ with the measure $\textrm{d}t/t$ and hence express the kinetic mixing parameter in terms of the moduli:
\begin{equation}
\chi \propto \frac{e^{\frac{2 \pi i}{U_2} \textrm{Im} \left( z (q + \bar{U} p) \right) }}{\frac{\pi}{U_2} {\left|q + \bar{U} p \right|}^2} .
\end{equation}
The proportionality factor is given by the group theoretical prefactors. This result is in fact independent of the K\"{a}hler modulus of the torus being wrapped. This is as expected based on holomorphy arguments: as a 1-loop correction, the kinetic mixing can depend only on the complex structure modulus as any dependence on the K\"ahler moduli is forbidden by a combination of holomorphy and shift symmetries.


\section{Kinetic mixing in general orbifolds}
\label{general}

Now that we have discovered kinetic mixing in the $\mathbb{T}^6/\mathbb{Z}_4$ orbifold, let us summarize the conditions for it to occur.
These conditions are applicable for models of this kind based on D3 branes on toroidal orbifolds.
We expect that they should be relaxed for Calabi-Yau models which have a more complicated topology.
\begin{enumerate}
\item
First of all we note that a non-anomalous U(1) defined on a single orbifold singularity cannot participate in mixing if it is massless at the same time, as both kinetic mixing and masses arise from the same coupling of the U(1) to a RR field. Thus, to obtain a non-anomalous and massless U(1) that is capable of producing kinetic mixing, we need at least two orbifold fixed points to define it on.
\item
In addition, it is crucial that the orbifold action allows for at least one $\mathcal{N}=2$ sector.  The presence of $\mathcal{N}=2$ sectors ensures the existence of further non-anomalous U(1)s beyond the diagonal $U(1)_{\mathrm{diag}}$ that have non-zero Green-Schwarz coupling and gain masses at the scale $M_s/R$. It is from these additional Abelian symmetry groups on different stacks of branes that we can build a massless U(1) that can participate in kinetic mixing.
\item
Furthermore, for two such U(1)s to exhibit kinetic mixing they have to be defined on singularities that must not all have the same separation in the compact space, otherwise the sums over winding modes will be equivalent and cancel one another.
\end{enumerate}
The last observation is consistent with previous results based on a general CFT computation employing vertex operators \cite{Durham}. There it was explained how anomaly-free U(1)s can mix kinetically while remaining massless when they are set up on stacks of branes that have different separations in the complex space. As mass mixing is induced by the exchange of massless closed string modes these effects can be cancelled among the different stacks of branes as massless modes are blind to the separations. Kinetic mixing, on the contrary, is also mediated by massive modes which do not necessarily cancel as they are sensitive to the geometry. This is the statement that the spectrum of massive winding
modes charged under both $U(1)$s is a function of the complex structure moduli.
This is exactly the behaviour that is observed in the $\mathbb{T}^6/\mathbb{Z}_4$ example presented above.

We now extend to more general models the conditions for kinetic mixing in orbifold models with D3 branes only,
and study further examples to illustrate the conclusions reached.


\subsection{General considerations}

A schematic calculation will enable us to examine whether kinetic mixing is possible in models with D3 branes at orbifold singularities when the U(1)s have no tree-level couplings. For simplicity we consider a single $\cN=2$ sector leaving one complex plane invariant. We define two massless Abelian groups ${\textrm{U}(1)}_a$ and ${\textrm{U}(1)}_b$ that are each linear combinations of U(1)s from multiple orbifold fixed points. We then assign charges $q_L=Q_a$ and $q_R = Q_b$ and expand the one-loop vacuum string amplitude to second order in the magnetic field $B$. In general $\mathrm{U}(1)_a$ and $\mathrm{U}(1)_b$ will share fixed points and hence both diagrams in figure \ref{fig:mixingdiagrams} can contribute to the amplitude. The contributions come solely from the $\mathcal{N}=2$ sector of the orbifold as the $\mathcal{N}=1$ sectors have to vanish to ensure tadpole cancellation and non-anomalous U(1)s.

\FIGURE[ht]{
\includegraphics[width=0.75\textwidth]{mixing.pdf}
\caption{String diagrams contributing to kinetic mixing of U(1)s defined accross multiple orbifold fixed points.}
\label{fig:mixingdiagrams}}
There are two different contributions to the vacuum amplitude: one comes from strings that start and end on the same stack of D3 branes; the second from strings that start and end at different orbifold fixed points and whose winding sums depend on the complex separation of the singularities in the compact space.  Schematically, the amplitude is:
\begin{equation}
{\mathcal{A}}_{\mathcal{N}=2}^{\mathcal{O}(B^2)} = \int \frac{\textrm{d}t}{t} \sum_{n,m} \left[
\sum\limits_{I} C_{I}\, e^{- \pi \frac{T_2}{U_2} {\left|U n + m \right|}^2 t} + \sum\limits_{I\neq J} C_{IJ}\, e^{- \pi \frac{T_2}{U_2} {\left|z_{IJ} + U n + m \right|}^2 t} \right],
\end{equation}
where the indices $I$ and $J$ run over orbifold fixed points contributing to the U(1)s and $z_{IJ}$ are their complex separations. The only dependence on the modular parameter $t$ arises from the winding modes as the string oscillator tower collapses to a single number in the $\mathcal{N}=2$ sector. The coefficients $C_I$ and $C_{IJ}$ are numerical prefactors that derive from the traces over the Chan-Paton states and are hence independent of $t$.

As we are interested in kinetic mixing between hidden U(1)s we need to ensure that there is no tree-level coupling between the two Abelian gauge bosons. In the string calculation this translates into the requirement that the above amplitude should vanish in the open string IR regime $t \rightarrow \infty$. In this limit massive strings do not occur and only strings localized at single fixed points contribute:
\begin{equation}
{\mathcal{A}}_{\mathcal{N}=2}^{\mathcal{O}(B^2)} \underset{t \rightarrow \infty}{\rightarrow} \int \frac{\textrm{d}t}{t}
\sum\limits_{I} C_{I} .
\end{equation}
The divergence indicates a tree-level coupling between the U(1)s and to remove it we have to arrange the numbers of fractional branes across singularities accordingly so that the coefficient $\sum_I C_I$ vansihes. This conclusion here is the same whether the U(1)s share a fixed point or not.

Furthermore, we must analyse the string amplitude in the open string UV limit $t \rightarrow 0$. In this regime the diagram with both generators on the same boundary is proportional to tadpoles whereas the diagram with the generators on opposite boundaries will depend on the Green-Schwarz couplings of the U(1)s with a partially twisted RR-field. As we have to cancel tadpoles to guarantee the consistency of the theory and we exclusively consider massless U(1)s, the string vacuum amplitude has to vanish for small $t$. After Poisson resumming the string amplitude and re-expressing it using the modular parameter $l=1/t$ we arrive at:
\begin{equation}
{\mathcal{A}}_{\mathcal{N}=2}^{\mathcal{O}(B^2)} \underset{l \rightarrow \infty}{\rightarrow} \int \textrm{d}l \frac{2}{T_2} \left[ \sum_I C_I + \sum_{I\neq J}C_{IJ} \right] .
\end{equation}

\ni We already know that for hidden U(1)s we must have $\sum_I C_I = 0$. For the above amplitude to vanish for large $l$ we are thus forced to arrange our model such that $\sum_{I\neq J}C_{IJ}=0$ for strings starting and ending on different stacks. The vacuum amplitude must therefore take the form
\begin{equation}
{\mathcal{A}}_{\mathcal{N}=2}^{\mathcal{O}(B^2)} = \int \frac{\textrm{d}t}{t} \, \sum_{n,m} \left[\,\sum\limits_{I\neq J} C_{IJ} \,e^{- \pi \frac{T_2}{U_2} {\left|z_{IJ} + U n + m \right|}^2 t} \right]
\end{equation}
where  $\sum_{I\neq J}C_{IJ}=0$. The construction of models with kinetic mixing between hidden and massless U(1)s now depends crucially on the magnitude of the complex separations $z_{IJ}$.

If the the complex separations $z_{IJ}$ all have the same magnitude then the winding sums are identical and the vacuum amplitude factorises: the condition  $\sum_{I\neq J}C_{IJ}=0$ then forces the amplitude to vanish identically. An example of this behaviour is the $\mathbb{Z}_6$ orbifold which is studied below. However, if some of the complex separations $z_{IJ}$ have different magnitudes then the winding sums do not cancel in the regime of intermediate $t$. The integral then gives a finite contribution to kinetic mixing  despite vanishing in the IR and UV limits. This is the behaviour we have already seen in the $\mathbb{Z}_4$ orbifold model. In conclusion we see that to construct models with kinetic mixing the orbifold must have at least one $\cN=2$ sector and the presence of orbifold fixed points not all equidistant from one another in agreement with our earlier intuition.


\subsection{$\mathbb{T}^6 / \mathbb{Z}_6$}

An example where we do not expect to be able to construct models with kinetic mixing between hidden massless U(1)s is the orbifold ${\mathbb{T}}^6/{\mathbb{Z}}_{6}$ which is shown in figure \ref{Z6}. The orbifold action is defined by $\theta = (1/6, 1/6, -1/3)$ and hence there is one $\mathcal{N}=2$ sector generated by ${\theta}^3$. Crystallographic restriction forces the two-tori of the compact space to be defined on the $\textrm{SU}(3)$ root lattice. The orbifold twist then possesses three fixed points
\begin{align}
A: \quad & \Big(0,0, 0\Big), \\
B: \quad & \Big(0,0, e^{\frac{\pi i}{6}}/ \sqrt{3}\Big), \\
C: \quad & \Big(0,0, i/ \sqrt{3}\Big),
\end{align}
where we can place stacks of D3 branes with a CP embedding
\begin{equation}
\gamma_{\theta} = \textrm{diag}({\mathbbm{1}}_{n_0}, {\alpha}^1 {\mathbbm{1}}_{n_1}, {\alpha}^2 {\mathbbm{1}}_{n_2}, \ \dots \ , {\alpha}^{N-1} {\mathbbm{1}}_{n_{N-1}} )
\end{equation}
where $\alpha = e^{2 \pi i /6}$.
Cancellation of $\cN=1$ twisted tadpoles enforces $$n_0=n_2=n_4 \ \textrm{and} \ n_1=n_3=n_5$$ at each orbifold fixed point separately and, although we will not do it explicitly, $\mathcal{N}=2$ tadpoles can be cancelled across the singularities if the numbers of fractional branes are chosen accordingly.

\FIGURE[ht]{
\includegraphics[width=0.75\textwidth]{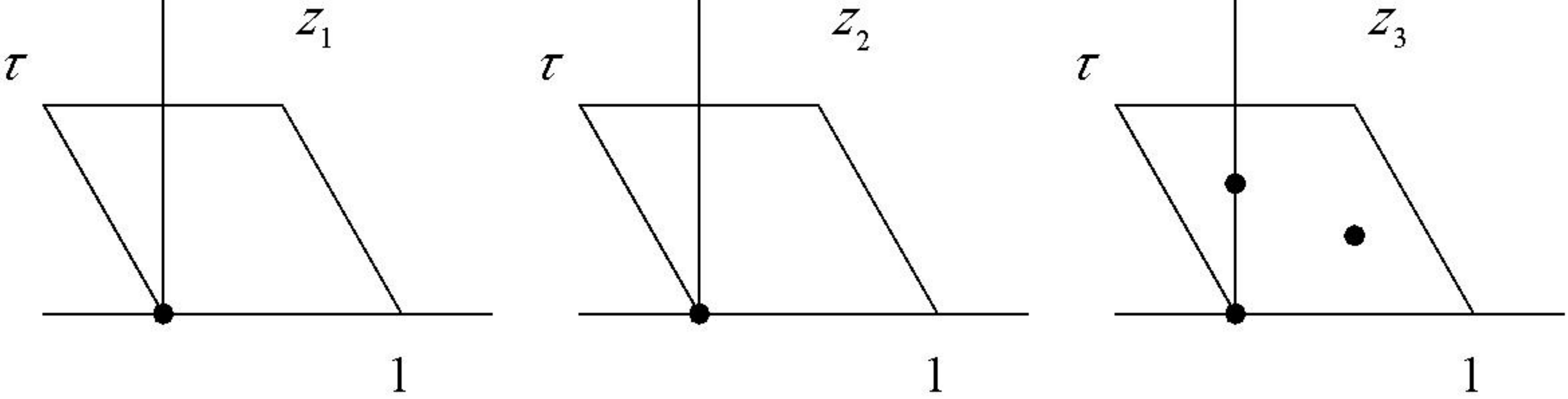}
\caption{The $\mathbb{T}^6/\mathbb{Z}_6$ orbifold. Dark circles correspond to fixed points under all orbifold twists and $\tau$ is the complex modulus of the tori.}
\label{Z6}}

As the ${\mathbb{Z}}_6$ orbifold exhibits one $\mathcal{N}=2$ sector we can find two mutually orthogonal non-anomalous U(1)s at each singularity: one is the diagonal U(1) which is massless and cannot participate in mixing, the second gains a mass due to a non-vanishing Green-Schwarz coupling to the $\cN=2$ RR two-form and is defined as
\begin{equation}
\textrm{U}(1)_{tw} =\frac{1}{{\mathcal{N}}_{tw}} \left[ \frac{1}{n_1}(\textrm{U}(1)_{0}+\textrm{U}(1)_{2}+\textrm{U}(1)_{4}) - \frac{1}{n_0}(\textrm{U}(1)_{1}+\textrm{U}(1)_{3}+\textrm{U}(1)_{5}) \right].
\end{equation}
where we were careful to include the normalisation such that $\textrm{Tr} \ Q_{tw}^2=1$. If we want to construct massless U(1)s that are capable of kinetic mixing we are again forced to define them over more than one orbifold fixed point. As the ${\mathbb{Z}}_6$ orbifold only possesses three fixed points these U(1)s inevitably share stacks of D3 branes. However, since the three orbifold fixed points are equidistant, then any amplitude contributing to kinetic mixing between massless hidden U(1)s must vanish.


\subsection{Multiple $\mathcal{N}=2$ sectors}
The effect of kinetic mixing is non-local from the viewpoint of the orbifold singularity as it relies on the propagation of $\mathcal{N}=2$ modes across the compact space. Hence it is imperative for the orbifold to exhibit at least one $\mathcal{N}=2$ sector.

So far we exclusively examined orbifold singularities that possess only one $\mathcal{N}=2$ sector. Within this framework we were not able to discover kinetic mixing between U(1)s that are defined on a single stack of D3-branes respectively. The reason for this failure is to be found in our condition on the U(1)s to be massless: To ensure massless Abelian gauge bosons we were forced to arrange for the trace factor to vanish in the $\mathcal{N}=2$ sector. As this trace also appears in the expression for the mixing parameter any kinetic mixing was thus automatically prohibited.

Our difficulties described above in constructing models that allow for kinetic mixing should be of interest for string model builders: A scenario of particular phenomenologogical relevance would be where one of the U(1)s involved in kinetic mixing is the weak hypercharge of the Standard Model of particle physics. In various string models the Standard Model is realised on a visible stack of branes
while there may also be hidden brane stacks. The hypercharge ${\textrm{U}(1)}_Y$ defined on the visible stack of branes can then mix kinetically with a second U(1) localised on a hidden stack leading to potentially observable effects \cite{Durham2, Durham}. Thus, from the point of view of model builders and string phenomenologists it would be interesting to find out whether kinetic mixing can occur at all between U(1)s defined on the worldvolume of a single stack of D3-branes.

We will now show that our observations regarding the impossibility of kinetic mixing between U(1)s on single stacks of branes generalise to the case of orbifolds with several $\mathcal{N}=2$ sectors. A general orbifold will have multiple $\mathcal{N}=2$ sectors which we can classify by the surfaces in the compact space which are kept constant by the geometrical action of the orbifold element in that sector. The $\mathcal{N}=2$ sectors that are distinct in this sense correspond to the various independent two-cycles in the compact space. Let us define a non-anomalous gauge group ${\textrm{U}(1)}_a$ at a single orbifold singularity as described earlier in this paper. In addition, we also ensure that ${\textrm{U}(1)}_a$ remains massless. To achieve this we have to arrange for all the Green-Schwarz (GS) couplings to vanish. The details of the calculation of Green-Schwarz couplings have already been described in this paper as well as in \cite{CP} and hence we will just sketch the results. In practice, we can get one Green-Schwarz coupling for each different $\mathcal{N}=2$ sector. The relevant string diagram is the annulus diagram with one generator $Q_a$ inserted at each boundary which we need to evaluate in the limit $t \rightarrow 0$. We get the following sum over $\mathcal{N}=2$ sectors:
\begin{equation}
{(\textrm{GS coupling})}^2 \propto \sum_{\substack{\textrm{distinct} \\ \mathcal{N}=2 \\ \textrm{sectors}}}{\left|\textrm{Tr} \left(Q_a {\gamma}_{\mathcal{N}=2}\right) \right|}^2 \prod_{i=1}^2 \left(-2 \sin \pi \theta_i \right) \int \frac{\textrm{d}t}{t^2} .
\end{equation}
For ${\textrm{U}(1)}_a$ to remain massless the above expression has to sum to zero. However, by examining each term closely we find that each $\mathcal{N}=2$ sector contributes to this sum by an amount geater or equal to zero. Obviously, the absolute value of the trace factor squared cannot be negative. In addition, given the constraint\footnote{The constraint $\theta_1 + \theta_2=1 \ \textrm{mod} \ 2$ arose when we applied a Riemann identity to simplify the contribution from string oscillators to the string amplitude. The identity is given in appendix \ref{Jacobi}.} $\theta_1 + \theta_2=1 \ \textrm{mod} \ 2$, the product
\begin{equation}
\prod_{i=1}^2 \left(-2 \sin \pi \theta_i \right) =4 \sin \pi \theta_1 \sin \pi \theta_2= 2[1+\cos(\pi \theta_1 - \pi \theta_2)]
\end{equation}
is non-negative. For this sum to be zero we need each summand to vanish. The only freedom we have to achieve this is to arrange for the trace factors to vanish:
\begin{equation}
\textrm{Tr} \left(Q_a {\gamma}_{\mathcal{N}=2}\right) =0 \quad \textrm{for all } \mathcal{N}=2 \textrm{ sectors.}
\end{equation}
Let us examine the consequences of this for kinetic mixing between this ${\textrm{U}(1)}_a$ and a hidden ${\textrm{U}(1)}_b$ defined on a different stack of branes. Again, the relevant string diagram is the annulus worldsheet with one generator $Q_a$ inserted on one boundary and the other $Q_b$ inserted on the other. Schematically, the amplitude can be written as:
\begin{equation}
\chi \propto \sum_{\substack{\textrm{distinct} \\ \mathcal{N}=2 \\ \textrm{sectors}}} \textrm{Re} \left\{\textrm{Tr} \left(Q_a {\gamma}_{\mathcal{N}=2}\right) \textrm{Tr} \left(Q_b {\gamma}_{\mathcal{N}=2}^{*}\right) \right\} \prod_{i=1}^2 \left(-2 \sin \pi \theta_i \right) \int \frac{\textrm{d}t}{t} \sum_{\substack{\textrm{winding}\\ \textrm{states}}} \exp \left( - m_{\substack{\textrm{winding}\\ \textrm{state}}}^2 \ t\right) .
\end{equation}
In this expression for the kinetic mixing parameter the same trace factors appear as in the calculation of Green-Schwarz couplings. Hence, if any one of the U(1)s is massless all the coresponding factors $\textrm{Tr} \left(Q {\gamma}_{\mathcal{N}=2}\right)$ are zero individually, thus prohibiting any kinetic mixing to occur.
From the considerations above we arrive at the following general statement: In toroidal orbifold models there is no kinetic mixing between massless U(1)s defined at single orbifold fixed points.


\subsection{$\mathbb{T}^6 /{\mathbb{Z}}_6^\prime$}
The ${\mathbb{Z}}_6^\prime$ orbifold action is $\theta = (1/6, 1/3, - 1/2)$ and there are twelve orbifold fixed points in the compact space in total. We use the same Chan-Paton-embedding as in the case of ${\mathbb{Z}}_6$. Sectors with twists $\theta^1$ and $\theta^5$ exhibit $\mathcal{N}=1$ supersymmetry and the cancellation of the associated tadpoles requires $n_0=n_2+n_3-n_5$ and $n_1=-n_2+n_4+n_5$. There are two distinct sectors with $\mathcal{N}=2$ supersymmetry generated by $\{\theta^2, \theta^4\}$ and $\{\theta^3 \}$ respectively: twists $\theta^2$ and $\theta^4$ both leave the third two-torus invariant while $\theta^3$ keeps the second two-torus fixed.

The presence of multiple $\mathcal{N}=2$ sectors has an interesting implication on the main objects of study in this work: we recall that for each orbifold element that leaves one torus fixed we get an additional non-anomalous U(1) at the orbifold singularity. As a result each ${\mathbb{Z}}_6^\prime$ supports up to three non-anomalous U(1)s at each orbifold singularity, one of which is the trivial diagonal combination encountered before. To find all non-anomalous U(1)s we solve equation (\ref{NAU1s}) for $r_k$ and insert the results into equation (\ref{cs}) to find the coefficients $c_j$ in the definition (\ref{Q}) of the non-anomalous U(1). The solutions for the coefficients are given by
\begin{align}
c_0 =& \frac{1}{6} \left(r_0 + r_2 + r_3 + r_4 \right), \\
c_1 =& \frac{1}{6} \left(r_0 - \frac{r_2}{2} - r_3 - \frac{r_4}{2} \right), \\
c_2 =& \frac{1}{6} \left(r_0  - \frac{r_2}{2} + r_3 - \frac{r_4}{2} \right), \\
c_3 =&  \frac{1}{6} \left(r_0 + r_2 - r_3 + r_4 \right), \\
c_4 =& c_2, \\
c_5 =& c_1,
\end{align}
where $r_0$,  $r_2$,  $r_3$ and  $r_4$ are arbitrary real numbers.\footnote{The variables $r_2$ and $r_4$ always appear in the same combination $r_2 + r_4$ and thus correspond to only one independent parameter. This is a consequence of $\theta^2$ and $\theta^4$ spanning the same $\mathcal{N}=2$ sector. Hence there are three independent parameters corresponding to three independent non-anomalous U(1)s.}

There are two possible Green-Schwarz couplings in the context of the ${\mathbb{Z}}_6^\prime$ singularity which can render non-anomalous U(1)s massive. The U(1)s can either couple to a closed string RR mode twisted by $\theta^3$ propagating across the second two-torus or to a RR mode twisted by $\{\theta^2, \theta^4 \}$ travelling across the third two-torus. Performing the string calculation of Green-Schwarz couplings we find that for U(1)s to remain massless in both $\mathcal{N}=2$ sectors we require
\begin{align}
\theta^2 + \theta^4: & \quad \textrm{Tr}(Q \gamma_{\theta^2})=\textrm{Tr}(Q \gamma_{\theta^4})=0 \ & \Rightarrow & \ r_2+r_4 =0, \\
\theta^3: & \quad \textrm{Tr}(Q \gamma_{\theta^3})=0 \ & \Rightarrow & \ r_3 =0,
\end{align}
where $Q$ is the generator of the U(1) in question. Given the results so far we define the following Abelian groups which span the whole space of non-anomalous U(1)s at a ${\mathbb{Z}}_6^\prime$ singularity:
\begin{eqnarray}
{\textrm{U}(1)}_{diag} &=& \frac{1}{{\mathcal{N}}_{diag}}\left(\frac{{\textrm{U}(1)}_0}{n_0} + \frac{{\textrm{U}(1)}_1}{n_1} + \frac{{\textrm{U}(1)}_2}{n_2} + \frac{{\textrm{U}(1)}_3}{n_3} + \frac{{\textrm{U}(1)}_4}{n_4} + \frac{{\textrm{U}(1)}_5}{n_5}\right),  \\
{\textrm{U}(1)}_{tw1} &=& \frac{1}{{\mathcal{N}}_{tw1}}\left(\frac{{\textrm{U}(1)}_0}{n_0} - \frac{{\textrm{U}(1)}_1}{n_1} + \frac{{\textrm{U}(1)}_2}{n_2} - \frac{{\textrm{U}(1)}_3}{n_3} + \frac{{\textrm{U}(1)}_4}{n_4} - \frac{{\textrm{U}(1)}_5}{n_5}\right), \\
{\textrm{U}(1)}_{tw2} &=& \frac{1}{{\mathcal{N}}_{tw2}}\left(2\frac{{\textrm{U}(1)}_0}{n_0} - \frac{{\textrm{U}(1)}_1}{n_1} - \frac{{\textrm{U}(1)}_2}{n_2} +2 \frac{{\textrm{U}(1)}_3}{n_3} - \frac{{\textrm{U}(1)}_4}{n_4} - \frac{{\textrm{U}(1)}_5}{n_5}\right).
\end{eqnarray}
The diagonal non-anomalous ${\textrm{U}(1)}_{diag}$ is defined by setting $r_2=r_3=r_4=0$ and it is massless in the global model. The other two non-anomalous U(1)s are massive: ${\textrm{U}(1)}_{tw1}$ is defined by $r_0=r_2=r_4=0$ and it gains a mass due to a coupling to the $\theta^3$-twisted RR mode; ${\textrm{U}(1)}_{tw2}$ is obtained by setting $r_0=r_3=0$ and it becomes massive by coupling to the $\theta^2$- and $\theta^4$-twisted RR mode. It is worth pointing out that we can state a basis for all non-anomalous U(1)s which, although not orthogonal, is given by U(1)s that at most show only one Green-Schwarz coupling.

\FIGURE[ht]{
\label{Z6prime}
\includegraphics[width=0.75\textwidth]{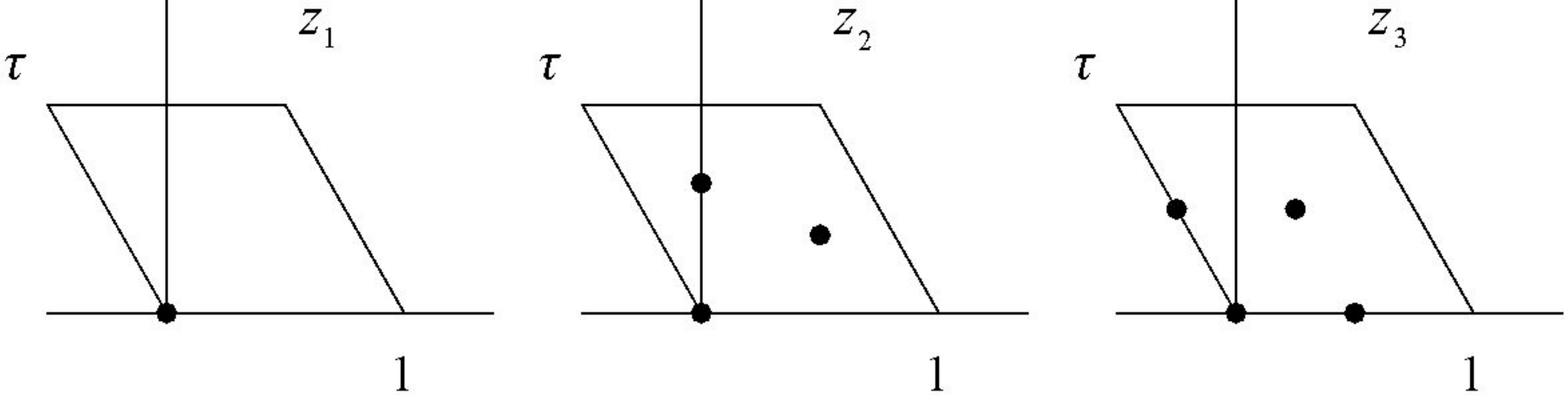}
\caption{The $\mathbb{T}^6/{\mathbb{Z}}_6^\prime$ orbifold. Dark circles correspond to fixed points under all orbifold twists and $\tau$ is the complex modulus of the tori.}}
We now turn to the study of kinetic mixing in this $\mathbb{T}^6/{\mathbb{Z}}_6^\prime$ orbifold model. Altogether there are twelve orbifold fixed points which are shown in figure. The positions of the fixed points resemble the configurations seen for ${\mathbb{Z}}_6$ in the second torus and for ${\mathbb{Z}}_4$ in the third torus. We will see that the phenomenology of kinetic mixing will indeed comprise both the effects seen in our ${\mathbb{Z}}_4$ and ${\mathbb{Z}}_6$ examples.

First of all we acknowledge that there cannot be any kinetic mixing between massless U(1)s defined at a single singularity. The general observation stated before is illustated in this specific example: The only massless U(1) possible on a single stack of branes is ${\textrm{U}(1)}_{diag}$ for which all trace factors vanish.

To arrange for kinetic mixing we are forced to define U(1)s that stetch over at least two orbifold fixed points. We will have to look for combinations of ${\textrm{U}(1)}_{tw1}$ and ${\textrm{U}(1)}_{tw2}$ which are massless. However, we are constrained by homology in the combinations that we are allowed to make, as each $\mathcal{N}=2$ sectors correspond to a two-cycle in the compact space. The two-cycle associated with $\theta^3$ connects all fixed points that only differ in the coordinate $z_2$ on the second two-torus.  The $\theta^2$- and $\theta^4$-sectors define a two-cycle which connects fixed points that only differ in the $z_3$-coordinate on the third two-torus. There are no two-cycles that connect fixed points that differ in both the $z_2$- and $z_3$-coordinate. Correspondingly our choices of massless combinations of ${\textrm{U}(1)}_{tw1}$ and ${\textrm{U}(1)}_{tw2}$ are limited. One option is to build a massless U(1) by arranging ${\textrm{U}(1)}_{tw1}$s across fixed points in the second two-torus. In this case the $\theta^2$- and $\theta^4$-sectors cannot contribute and $\theta^3$ spans the only relevant $\mathcal{N}=2$ sector. This leads to a situation which is identical with the setup encountered before in the $\mathbb{Z}_6$ orbifold. We can refer the reader to our analysis of the $\mathbb{Z}_6$ orbifold and conclude that no kinetic mixing will be observed in this case. Another possibility of assembling a massless U(1) is given by allocating ${\textrm{U}(1)}_{tw2}$s at fixed points that only differ in the $z_3$-coordinate. Here, the $\theta^3$ does not contribute and the relevant $\mathcal{N}=2$ sector is spanned by $\theta^2$ and $\theta^4$. We arrive at a setup which resembles the case encountered in the $\mathbb{Z}_4$ example. Again, all conditions for kinetic mixing are fulfilled as in the case of the $\mathbb{Z}_4$ orbifold and we can record that the $\mathbb{T}^6/{\mathbb{Z}}_6^\prime$ orbifold indeed allows for kinetic mixing in the same way as was discovered in the $\mathbb{Z}_4$ example.

Having established kinetic mixing in the case of the ${\mathbb{Z}}_6^\prime$ orbifold we observe that it obeys all the principles that we discovered in the $\mathbb{Z}_4$ example.


\subsection{$\mathbb{T}^6 / \Delta_{27}$}
As a final example we examine the $\Delta_{27}$ singularity which was studied in \cite{C, Delta271, Delta272, Delta273}. It is a non-Abelian singularity which also exists as a particular case of the $dP_8$ singularity \cite{dP}. $\Delta_{27}$ is the finite non-Abelian subgroup of SU(3) generated by
\begin{align}
e_1 : & (z_1, z_2, z_3) \rightarrow (\omega z_1, \omega^2 z_2, z_3), \\
e_2 : & (z_1, z_2, z_3) \rightarrow (z_1, \omega z_2,\omega^2 z_3), \\
e_3 : & (z_1, z_2, z_3) \rightarrow (z_3, z_1, z_2),
\end{align}
where $\omega = \exp (2 \pi i/3)$. The generators satisfy
\begin{equation}
e_1^3=e_2^3=e_3^3=1, \quad e_1e_2=e_2e_1, \quad e_3e_1=e_2e_3, \quad e_3e_2=e_1^2e_2^2e_3.
\end{equation}
The 27 elements of the group can be written as $e_1^{\alpha}e_2^{\beta}e_3^{\gamma}$ with $\alpha, \ \beta, \ \gamma = 1 \dots 3$.
The eleven conjugacy classes are
\begin{align}
\nonumber & \{1\}, \{e_1e_2^2\}, \{e_1^2e_2\}, \{e_1, e_2, e_1^2 e_2^2\}, \{e_1^2, e_2^2, e_1 e_2\}, \{e_3, e_1e_2^2e_3, e_1^2e_2e_3\}, \{e_1e_3, e_2e_3, e_1^2e_2^2e_3\},\\
& \{e_1^2e_3, e_2^2e_3, e_1e_2e_3\}, \{e_3^2, e_1e_2^2e_3^2, e_1^2e_2e_3^2\}, \{e_1e_3^2, e_2e_3^2, e_1^2e_2^2e_3^2\}, \{e_1^2e_3^2, e_2^2e_3^2, e_1e_2e_3^2\} .
\end{align}
Corresponding to these are eleven irreducible representations, $\mathbf{3} + \mathbf{3^{*}} + 9 \times \mathbf{1}$. The nine one-dimensional irreducible representations are given by
\begin{equation}
\gamma_{e_1}=\gamma_{e_2}= \omega^{\alpha}, \quad \gamma_{e_3}=\omega^{\beta},
\end{equation}
with $\alpha, \ \beta =0,1,2$. The three-dimensional irreducible representations are given by the defining representation and its complex conjugate:
\begin{align}
&\gamma_{e_1}^{\mathbf{3}}= \left( \begin{array}{ccc} \omega & 0 & 0 \\ 0 & \omega^2 & 0 \\ 0 & 0 & 1 \end{array} \right), \ \gamma_{e_2}^{\mathbf{3}}= \left( \begin{array}{ccc} 1 & 0 & 0 \\ 0 & \omega & 0 \\ 0 & 0 & \omega^2 \end{array} \right), \
\gamma_{e_3}^{\mathbf{3}}= \left( \begin{array}{ccc} 0 & 0 & 1 \\ 1 & 0 & 0 \\ 0 & 1 & 0 \end{array} \right), \\
&\gamma_{e_1}^{\mathbf{3^*}}= \left( \begin{array}{ccc} \omega^2 & 0 & 0 \\ 0 & \omega & 0 \\ 0 & 0 & 1 \end{array} \right), \ \gamma_{e_2}^{\mathbf{3^*}}= \left( \begin{array}{ccc} 1 & 0 & 0 \\ 0 & \omega^2 & 0 \\ 0 & 0 & \omega \end{array} \right), \
\gamma_{e_3}^{\mathbf{3^*}}= \left( \begin{array}{ccc} 0 & 0 & 1 \\ 1 & 0 & 0 \\ 0 & 1 & 0 \end{array} \right).
\end{align}

The action on the CP degrees of freedom can be decomposed into a direct sum over the irreducible representations:
\begin{equation}
\left({\bigoplus}_{i=1}^9 n_i \times {\mathbf{1}}_i\right) \oplus (n_{10} \times \mathbf{3}) \oplus (n_{11} \times \mathbf{3^*}).
\end{equation}
\FIGURE[ht]{
\label{Quiver27}
\includegraphics[width=0.75\textwidth]{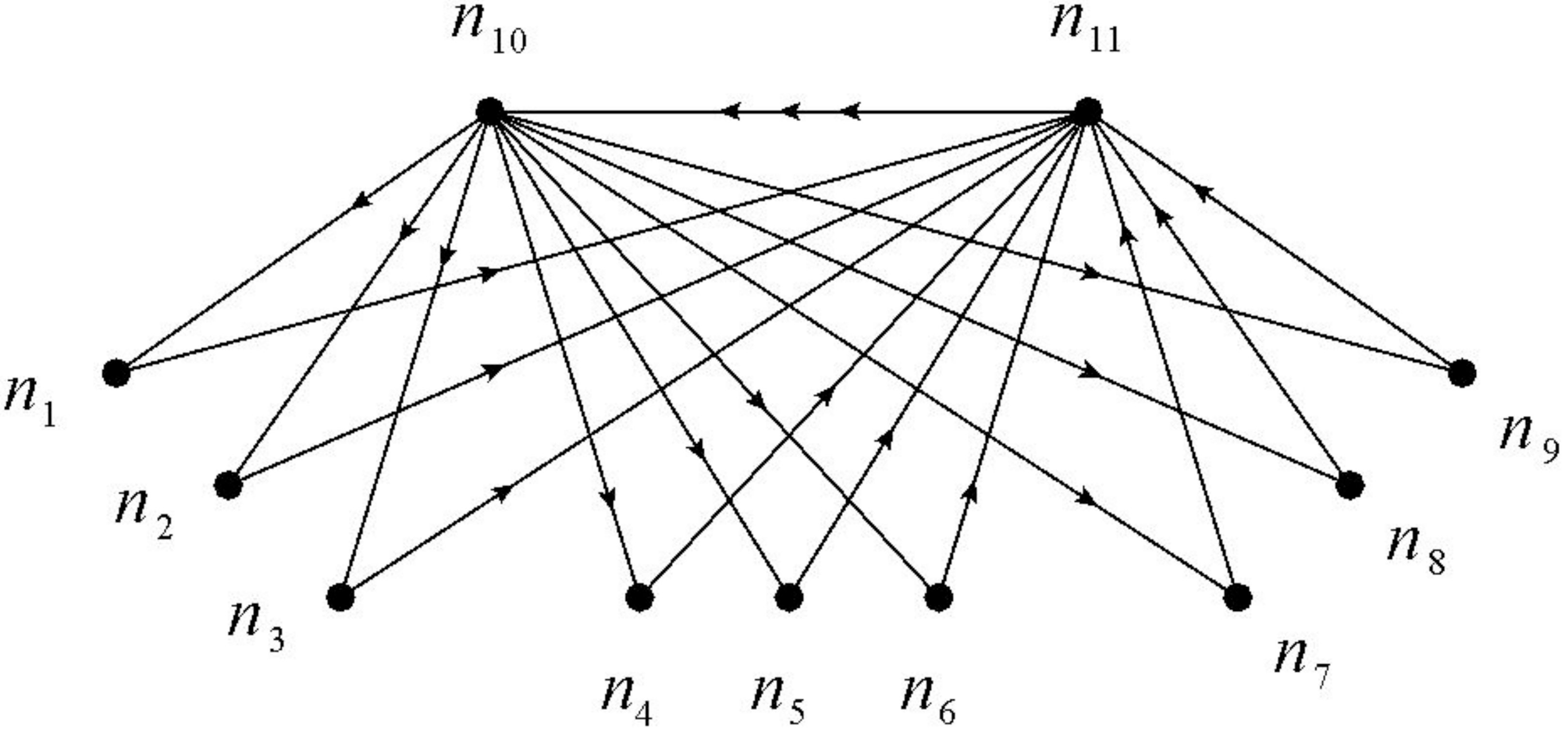}
\caption{The quiver for the $\Delta_{27}$ singularity}
}
We arrive at a gauge theory on the D3-branes which is depicted by the quiver in figure \ref{Quiver27}. Consistency requires
\begin{equation}
n_{10}=n_{11} =\frac{n_1 + n_2 + n_3 + n_4 + n_5 + n_6 + n_7 + n_8 + n_9}{3}.
\end{equation}

In the orbifold projection the identity group element preserves $\mathcal{N}=4$ supersymmetry as usual and the two elements $e_1e_2^2$ and $e_1^2e_2$ are $\mathcal{N}=1$ sectors. The remaining 24 group elements allow for $\mathcal{N}=2$ supersymmetry and it is these $\mathcal{N}=2$ sectors that we will be mostly interested in. In terms of the complex planes that are kept fixed by the geometrical action, there are
12 possibilities among the $\mathcal{N}=2$ sectors:\footnote{Note though that the number of twisted modes is determined by the number
of $\mc{N}=2$ conjugacy classes, which is nine.}
\begin{align}
\nonumber \mathcal{N}=2 \textrm{ group element} & \quad \textrm{fixed plane } (z_1, z_2, z_3) \\
\nonumber e_1, \ e_1^2: & \quad (0,0,z) \\
\nonumber e_2, \ e_2^2: & \quad (z,0,0) \\
\nonumber e_1e_2, \ e_1^2e_2^2: & \quad (0,z,0) \\
\nonumber e_3, \ e_3^2: & \quad (z,z,z) \\
\nonumber e_1e_3, \ e_1e_2e_3^2: & \quad (\omega z,z,z) \\
\nonumber e_2e_3, \ e_1^2e_3^2: & \quad (z, \omega z,z) \\
\nonumber e_1^2e_2^2e_3, e_2^2e_3^2: & \quad (\omega^2 z, \omega^2 z, z) \\
\nonumber e_1e_2e_3, \ e_2e_3^2: & \quad (\omega z, \omega z, z) \\
\nonumber e_1^2e_3, \ e_1^2e_2^2e_3^2: & \quad (\omega^2 z, z,z) \\
\nonumber e_2^2e_3, \ e_1e_3^2: & \quad (z, \omega^2 z, z) \\
\nonumber e_1e_2^2e_3, \ e_1^2e_2e_3^2: & \quad (\omega z, \omega^2 z, z) \\
e_1^2e_2e_3, \ e_1e_2^2e_3^2: & \quad (\omega^2 z, \omega z, z)
\end{align}

We now consider a global model from compactifying on $\mathbb{T}^6/\Delta_{27}$. The compact space has exactly three fixed points
\begin{equation}
\Big(0,0,0\Big) , \quad
\Big(\frac{e^{\pi i/6}}{\sqrt{3}}, \frac{e^{\pi i/6}}{\sqrt{3}}, \frac{e^{\pi i/6}}{\sqrt{3}}\Big) , \quad
\Big(\frac{i}{\sqrt{3}}, \frac{i}{\sqrt{3}}, \frac{i}{\sqrt{3}}\Big),
\end{equation}
which are illustrated in figure \ref{Delta27}
\FIGURE[ht]{
\label{Delta27}
\includegraphics[width=0.75\textwidth]{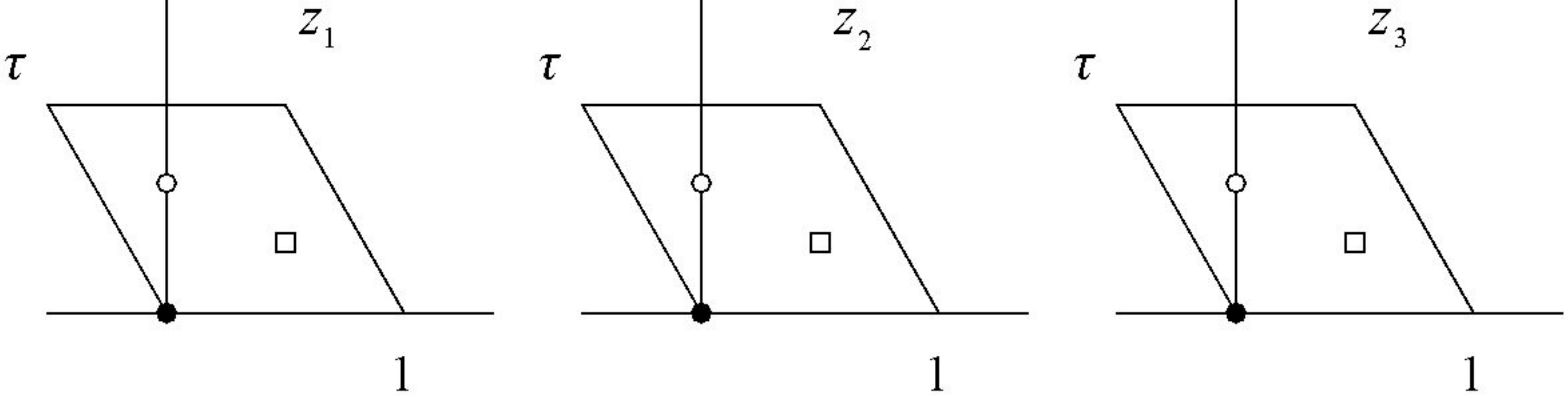}
\caption{The $\mathbb{T}^6/\Delta_{27}$ orbifold. The filled circle, the hollow circle and the square denote the coordinates of the three fixed points in the compact space.}}
We will now examine whether kinetic mixing between massless U(1)s can occur in the $\mathbb{T}^6/\Delta_{27}$ orbifold. The presence of the different $\mathcal{N}=2$ sectors guarantees the existence of non-anomalous U(1)s beyond the trivial diagonal combination. Out of these we can define massless U(1)s shared by at least two orbifold singularities which, in principle, could then mix kinetically. However, we need not to go into much detail to see that this will not be possible. Kinetic mixing is, as usual, communicated by winding states connecting the various orbifold fixed points. By studying the various fixed planes of the $\mathcal{N}=2$ sectors we see that the fixed points can be reached by winding states in all $\mathcal{N}=2$ sectors except in the ones spanned by $\{e_1,e_1^2\}$, $\{e_2, e_2^2\}$ and $\{e_1e_2, e_1^2e_2^2 \}$. Nevertheless, as the orbifold fixed points are equidistant to one another in the compact space, the sums over winding modes will be equivalent in all $\mathcal{N}=2$ sectors. We can then factor out the winding modes and are left with an expression that vanishes due the U(1)s being massless. It is the same principle that prohibits kinetic mixing as the discovered in the case of the $\mathbb{Z}_6$ orbifold.


\section{Conclusion}
We endeavour to discover kinetic mixing between massless U(1)s in toroidal orbifold models with fractional D3-branes at orbifold singularities. Calculations are made from a string theoretical perspective employing the background field method. In the course of the work we reaffirm results derived in \cite{Durham} but also present specific string models which we are able to examine in the search for kinetic mixing. In particular, we show that kinetic mixing is possible for some orbifolds and explore kinetic mixing in the example of the $\mathbb{Z}_4$ orbifold in detail. There we computed the size of 1-loop kinetic mixing and found it is a function of the complex structure of the torus with typical magnitude in the range
$10^{-1} -10^{-3}$.

Further, we are able to establish general rules regarding the possibility of kinetic mixing for D3 brane models in toriodal orbifolds:
Kinetic mixing is a non-local effect with respect to the orbifold singularity and it depends on the exchange of $\mathcal{N}=2$ RR modes across the compact space. Hence it is imperative for the orbifold to display at least one $\mathcal{N}=2$ sector. In addition, for each distinct $\mathcal{N}=2$ sector the orbifold singularity supports one additional non-anomalous U(1) which can then participate in kinetic mixing.

One important result is that U(1)s which are defined on a single stack of branes each are unable to mix kinetically if massless at the same time. This grounds on kinetic mixing and mass mixing originating from the same string diagrams as both kinetic and mass mixing rely on the exchange of $\mathcal{N}=2$ RR modes. The only difference is that in the case of kinetic mixing winding modes need to be included whereas mass mixing is evaluated in a limit where the winding modes collapse into a universal factor. In the case of U(1)s that are only supported at a single stack of branes the winding modes are universal and factor out parallelling the behaviour of mass mixing. In this scenario kinetic mixing is observed to vanish automatically for massless U(1)s as predicted in \cite{Durham}.

To arrange for kinetic mixing in toroidal orbifolds we are thus forced to consider U(1)s that are defined over at least two orbifold singularities. These U(1)s can be constructed such as to be massless while simultaneously displaying kinetic mixing. For this phenomenon to occur it is crucial that the toroidal orbifold possesses fixed points which are not equidistant to one another in the compact space. Strings stretched between fixed points that have different separations in the compact space result in distinct expressions for the winding modes. While we can enforce that mass mixing vanishes between the two U(1)s, the same cancellation is prevented in the kinetic mixing calculation due to the different winding modes. We examined this scenario in detail for the toroidal $\mathbb{Z}_4$ orbifold and showed that kinetic mixing can also occur in the case of $\mathbb{Z}_6^{\prime}$ models. We also established the non-existence of kinetic mixing in models based on the $\mathbb{Z}_6$ and the non-Abelian $\Delta_{27}$ orbifolds which we attribute to the orbifold singularities being equidistant in the toroidal compact space. This behaviour is again consistent with general predictions made in \cite{Durham}.

Besides, we find that the kinetic mixing parameter, while depending on the complex moduli of the compact space, does not depend on the
 K\"{a}hler moduli. This is a consequence of the holomorphy properties of the gauge kinetic function, which forbid the K\"ahler moduli
 appearing at the 1-loop level. As a result the magnitude of kinetic mixing is independent of the size of the compact space.

Our analysis was limited to models with D3-branes located at singularities in an orbifolded toroidal space.
We expect our conditions on kinetic mixing to be relaxed in Calabi-Yau models with more general topologies that
that of toroidal orbifolds.
Other natural extensions would be to include D7 branes that wrap two of the tori in the compact space
 or to consider local orientifold models. It would be interesting to see how our results for D3 models are modified in these cases.

\acknowledgments{
We thank Mark Goodsell and Eran Palti for discussions.
JC is funded by a Royal Society University Research Fellowship and by Balliol College, Oxford. MB and LW are funded by the Science and Technology Facilities Council. }


\appendix
\section{Gauge threshold corrections for local $\mathbb{Z}_M \times \mathbb{Z}_N$ orbifolds}
\label{ZMZN}
The methods used to calculate kinetic mixing are closely related to the techniques involved in examining gauge threshold corrections. Hence we will seize the opportunity to examine gauge threshold corrections for models that have not been included in the previous paper on this topic \cite{C}. This appendix can thus be regarded as an extension of this previous work, however, as the the calculational methods are very similar to the techniques used in the main text, it is not out of place to present the results here.

We will examine local models based on D3-branes located at a $\mathbb{C}^3/(\mathbb{Z}_M \times \mathbb{Z}_N)$ orbifold singularity. These models have already been studied for the case that discrete torsion is present \cite{DouglasZMZN, Berenstein}. Orientifolds with orbifold group $\mathbb{Z}_M \times \mathbb{Z}_N$ are treated in \cite{Zwart}.

We examine the case where discrete torsion is absent. The orbifold group is spanned by the two generators $\theta$ and $\eta$ corresponding to the twists by $\mathbb{Z}_M$ and $\mathbb{Z}_N$ respectively.  As we wish to obtain a low energy spectrum that preserves $\mathcal{N}=1$ supersymmetry we choose the following action for the group generators:
\begin{align}
\theta:& \ (z_1, z_2, z_3) \rightarrow (\alpha z_1, \alpha^{-1} z_2, z_3), \\
\eta:& \ (z_1, z_2, z_3) \rightarrow ( z_1, \beta z_2,\beta^{-1} z_3),
\end{align}
where $\alpha= \exp(2\pi i/M)$ and $\beta=\exp(2\pi i/N)$. The orbifold group consists of all the elements $\theta^k \eta^l$ with $k= 0 \dots M-1$ and $l= 0 \dots N-1$. The group $\mathbb{Z}_M \times \mathbb{Z}_N$ possesses $M \cdot N$ irreducible representations which are given by $\alpha^k \beta^l$ with $k= 0 \dots M-1$ and $l= 0 \dots N-1$ as before. As usual, we need to embed the orbifold action in the gauge group by stating its effect on the Chan-Paton degrees of freedom: The matrices $\gamma_{\theta,3}$ and $\gamma_{\eta,3}$ are block diagonal where each block corresponds to an irreducible representation of $\mathbb{Z}_M \times \mathbb{Z}_N$:
\[
\gamma_{\theta,3} =
 \begin{pmatrix}
 \textrm{diag}(\mathbbm{1}_{n_0}, \alpha \mathbbm{1}_{n_1}, \dots, \alpha^{M-1}\mathbbm{1}_{n_{M-1}}) & 0 & \cdots & 0 \\
  0 & \textrm{diag}(\mathbbm{1}_{n_M}, \alpha \mathbbm{1}_{n_{M+1}}, \dots, \alpha^{M-1}\mathbbm{1}_{n_{2M-1}}) & \cdots & 0 \\
  \vdots  & \vdots  & \ddots & \vdots  \\
  0 & 0 & \cdots & \textrm{diag}(\dots)
 \end{pmatrix},
\]
\[
\gamma_{\eta,3} =
 \begin{pmatrix}
 \textrm{diag}(\mathbbm{1}_{n_0}, \mathbbm{1}_{n_1}, \dots, \mathbbm{1}_{n_{M-1}}) & 0 & \cdots & 0 \\
  0 & \textrm{diag}(\beta \mathbbm{1}_{n_M}, \beta \mathbbm{1}_{n_{M+1}}, \dots, \beta \mathbbm{1}_{n_{2M-1}}\dots) & \cdots & 0 \\
  \vdots  & \vdots  & \ddots & \vdots  \\
  0 & 0 & \cdots & \textrm{diag}(\dots)
 \end{pmatrix}.
\]
Thus, for each group element $\theta^k \eta^l$,  we can give the corresponding CP action $\gamma_{\theta,3}^k \gamma_{\eta,3}^l$. After calculating the spectrum using standard techniques one arrives at a $\prod_{i=0}^{MN-1} \textrm{U}(n_i)$ gauge theory with bifundamental matter.

Before we calculate gauge threshold corrections we need to ensure that we are working with a consistent model. This requires the cancellation of all $\mathcal{N}=1$ tadpoles at the singularity and is described in the main text. In particular, we have to satisfy the condition set by equation (\ref{N1tadpoles}). As we will only be interested in the local model we do not specify the global completion and hence do not need to cancel $\mathcal{N}=2$ tadpoles.
The threshold calculation can be sketched as follows: To extract information about the $\beta$-function for the gauge group $\textrm{SU}(n_a)$ we embed the string endpoint charges $q_L=-q_R$ within the gauge group in question and analyse the one-loop vacuum string amplitude at order $\mathcal{O}(B^2)$:
\begin{equation}
\label{threshold}
\mathcal{A}^{\mathcal{O}(B^2)} = \frac{1}{2} {\left(\frac{B}{2 \pi^2} \right)}^2 \int_0^{\infty} \frac{\textrm{d}t}{8t} \Delta_a(t) .
\end{equation}
In the open string IR limit $t \rightarrow \infty$, the integrand approaches the field theory $\beta$-function coefficient $\Delta_a \rightarrow b_a$. The stringy physics is encoded in the UV limit $t \rightarrow 0$. In a consistent compact model the integral in  (\ref{threshold}) will be finite in the UV as, for non-Abelian groups, $\Delta_a(t)$ vanishes for small $t$ due to tadpole cancellation. The threshold corrections are encoded in the exact behaviour of $\Delta_a$ for $t \rightarrow 0$. In previous work it was found that the running of the gauge coupling is different in distinct orbifold sectors depending on the amount of supersymmetry preserved \cite{C, CP}. Schematically, the results are
\begin{equation}
  \Delta_a^{(k)} = \left\{
  \begin{array}{l l}
    b_a^{(k)} \ \Theta \left[t-\frac{1}{M_s^2}\right] + \textrm{small} & \quad \mathcal{N}=1 \textrm{ sector}\\
    b_a^{(k)} \ \Theta \left[t-\frac{1}{{(RM_s)}^2}\right] + \textrm{small} & \quad \mathcal{N}=2 \textrm{ sector}\\
    0 & \quad \mathcal{N}=4 \textrm{ sector}
  \end{array} \right.
\end{equation}
where $\Theta$ is the Heaviside theta function and $R$ is the bulk radius. In the following we will show that this form of threshold corrections also arises in local models at $\mathbb{Z}_M \times \mathbb{Z}_N$ orbifold singularities. To be specific, we will be considering two examples.


\subsection{$\mathbb{Z}_2 \times \mathbb{Z}_4$ orbifold singularity}

The $\mathbb{Z}_2 \times \mathbb{Z}_4$ singularity is generated by the actions $\theta=\frac{1}{2}(1,-1,0)$ and $\eta=\frac{1}{4}(0,1,-1)$.
\FIGURE[ht]{
\label{Z2xZ4}
\includegraphics[width=0.75\textwidth]{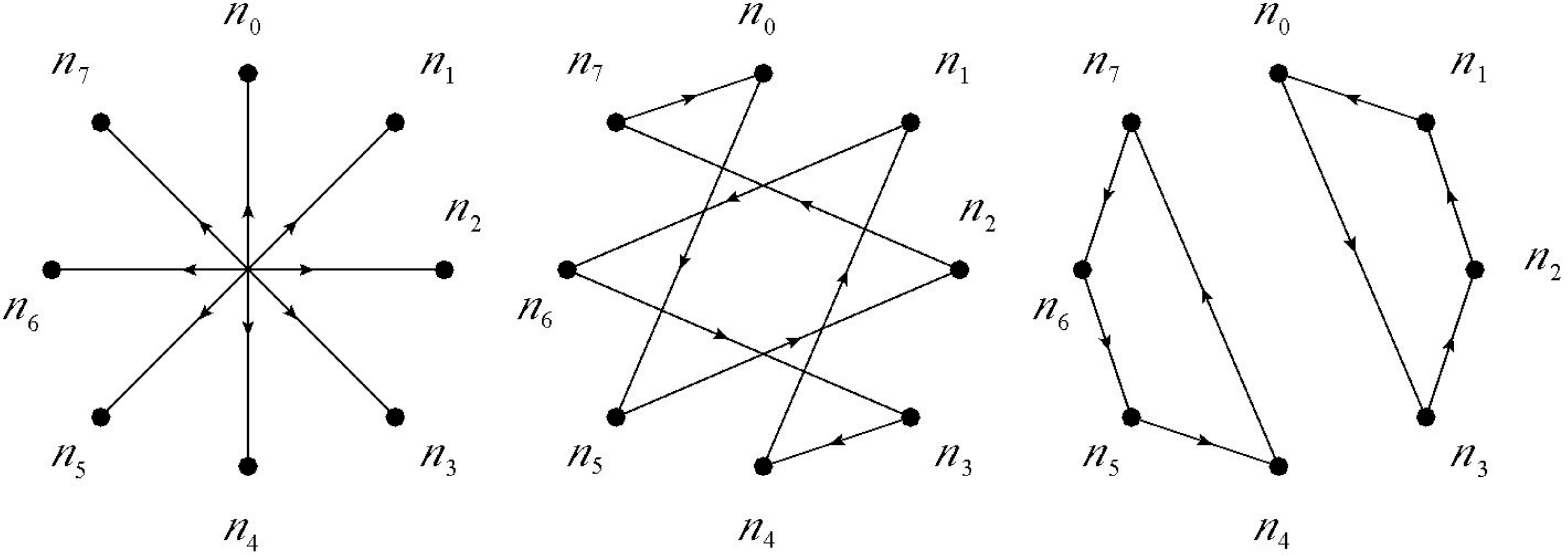}
\caption{Quiver for the $\mathbb{Z}_2 \times \mathbb{Z}_4$ orbifold singularity. For better visualisation the quiver is split into three diagrams. The diagrams show the matter content that is allowed by the orbifold twists $\{\theta_1, \ \eta_1\}$, $\{\theta_2, \ \eta_2\}$  and $\{\theta_3, \ \eta_3\}$ respectively.}}
The low energy spectrum is shown in the quiver diagram \ref{Z2xZ4}. There are two $\mathcal{N}=1$ sectors generated by $\{ \theta \eta, \theta \eta^3 \}$ and five $\mathcal{N}=2$ sectors $\{ \theta, \theta \eta^2, \eta, \eta^2, \eta^3 \}$.
Cancellation of twisted tadpoles requires $n_0 = n_2 + n_4 - n_6$ and $n_1 = n_3 + n_5 - n_7$.
Thresholds are calculated by assigning $q_L = - q_R = \textrm{diag}(Q_{\textrm{SU}(n_0)}, 0, \dots, 0)$ where $Q_{\textrm{SU}(n_0)}= \frac{1}{\sqrt{8}}(1,-1,0,\dots,0)$. Performing the calculation according to \cite {C, CP} gives:
\begin{align}
\mathcal{N}=1:& \quad \theta \eta + \theta \eta^3:&  \ {\left(\frac{B}{2 \pi^2}\right)}^2\int \frac{\textrm{d}t}{8t} (-1)& \left[n_0 -n_2 - n_4 + n_6 \right]  \\
\mathcal{N}=2:& \quad \eta + \eta^3:&  \ {\left(\frac{B}{2 \pi^2}\right)}^2\int \frac{\textrm{d}t}{8t} \frac{(-1)}{2}& \left[n_0 -n_2 + n_4 - n_6 \right] \\
\mathcal{N}=2:& \quad \eta^2:&  \ {\left(\frac{B}{2 \pi^2}\right)}^2\int \frac{\textrm{d}t}{8t} \frac{(-1)}{2}& \left[n_0 -n_1 +n_2 -n_3 + n_4 -n_5 + n_6 -n_7 \right] \\
\mathcal{N}=2:& \quad \theta:&  \ {\left(\frac{B}{2 \pi^2}\right)}^2\int \frac{\textrm{d}t}{8t} \frac{(-1)}{2}& \left[n_0 + n_1 +n_2 +n_3 - n_4 -n_5 - n_6 -n_7 \right] \\
\mathcal{N}=2:& \quad \theta \eta^2:&  \ {\left(\frac{B}{2 \pi^2}\right)}^2\int \frac{\textrm{d}t}{8t} \frac{(-1)}{2}& \left[n_0 - n_1 +n_2 -n_3 - n_4 +n_5 - n_6 +n_7 \right]
\end{align}
Combining all sectors gives the correct $\beta$-function coefficient for $\textrm{SU}(n_0)$:
\begin{equation}
b_0= -3n_0 + \frac{1}{2} (n_1 + n_3 +2n_4 + n_5 + n_7)
\end{equation}
We find that the contribution from $\mathcal{N}=1$ sectors vanishes once tadpole cancellation is imposed. The threshold corrections are hence sourced entirely by the $\mathcal{N}=2$ sectors which, in the local model, give divergent contributions in the open string UV limit. In a global model these divergences are cured once winding modes are included that explore the whole of the compact space. Thus we conclude that in the $\mathcal{N}=2$ sector the field theoretical running of the gauge coupling is observed up to bulk winding scale. This is the exact behaviour that has been observed in previous work on $\mathbb{Z}_N$ orbifold \cite{C} and orientifold singularities \cite{CP}.


\subsection{$\mathbb{Z}_3 \times \mathbb{Z}_3$ orbifold singularity}

We conclude this section with a further example that will confirm the above observations. The generators of the $\mathbb{Z}_3 \times \mathbb{Z}_3$ singularity are $\theta=\frac{1}{3}(1,-1,0)$ and $\eta=\frac{1}{3}(0,1,-1)$.
\FIGURE[ht]{
\label{Z3xZ3}
\includegraphics[width=0.75\textwidth]{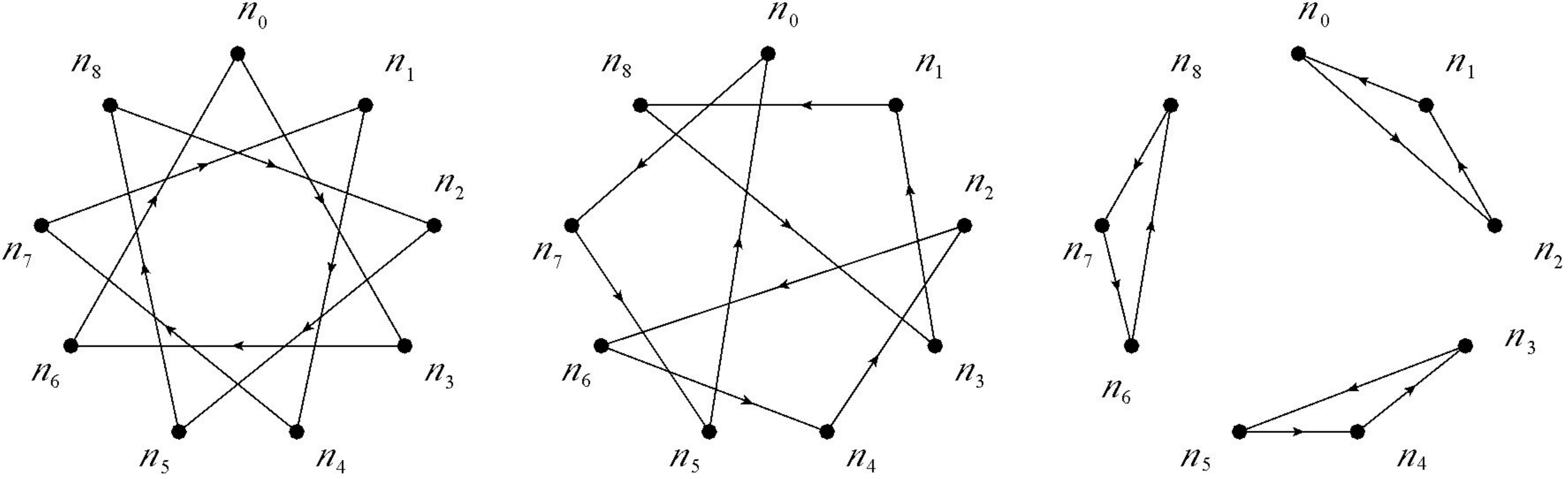}
\caption{Quiver for the $\mathbb{Z}_3 \times \mathbb{Z}_3$ orbifold singularity. For better visualisation the quiver is split into three diagrams. The diagrams show the matter content that is allowed by the orbifold twists $\{\theta_1, \ \eta_1\}$, $\{\theta_2, \ \eta_2\}$  and $\{\theta_3, \ \eta_3\}$ respectively.}
}
The quiver is shown in figure \ref{Z3xZ3}. The vanishing of twisted tadpoles requires:
\begin{align}
n_0 =& -n_4 -n_8 + \frac{1}{2} (n_1 + n_2 + n_3 + n_5 + n_6 + n_7) \\
n_1=& n_2 + n_3 -n_5 - n_6 + n_7 .
\end{align}
The two sectors that preserve $\mathcal{N}=1$ supersymmetry are spanned by $\{ \theta^2 \eta, \theta \eta^2 \}$ while there are six $\mathcal{N}=2$ sectors $\{ \theta, \theta^2, \eta, \eta^2, \theta \eta, \theta^2 \eta^2 \}$. Calculating the threshold corrections gives:
\begin{align}
\theta \eta^2 + \theta^2 \eta:&  \ {\left(\frac{B}{2 \pi^2}\right)}^2\int \frac{\textrm{d}t}{8t} \frac{(-1)}{2} \left[2n_0 -n_1 -n_2 -n_3 +2 n_4 -n_5- n_6 -n_7 +2 n_8 \right]  \\
\theta + \theta^2:&  \ {\left(\frac{B}{2 \pi^2}\right)}^2\int \frac{\textrm{d}t}{8t} \frac{(-1)}{3} \left[2n_0 +2n_1 +2n_2 -n_3 - n_4 -n_5 - n_6 -n_7 -n_8 \right] \\
\eta+\eta^2:&  \ {\left(\frac{B}{2 \pi^2}\right)}^2\int \frac{\textrm{d}t}{8t} \frac{(-1)}{3} \left[2n_0 -n_1 -n_2 +2n_3 - n_4 -n_5 +2 n_6 -n_7 -n_8\right] \\
\theta \eta + \theta^2 \eta^2:&  \ {\left(\frac{B}{2 \pi^2}\right)}^2\int \frac{\textrm{d}t}{8t} \frac{(-1)}{3} \left[2n_0 - n_1 -n_2 -n_3 - n_4 +2n_5 - n_6 +2n_7-n_8 \right] .
\end{align}
Summing over the individual sectors once more results in the correct $\beta$-function coefficient:
\begin{equation}
b_0= -3n_0 +\frac{1}{2} (n_1 +n_2 + n_3 + n_5 + n_6 + n_7) .
\end{equation}
The interpretation of the result coincides with the discussion given above. Contributions from $\mathcal{N}=1$ sectors vanish once anomaly cancellation is enforced and threshold corrections arise solely from $\mathcal{N}=2$ sectors permitting field theory gauge running up to the bulk winding scale.


\section{Properties of Jacobi theta functions}
\label{Jacobi}
In this section we will summarize definitions and identities related to Jacobi theta functions. We denote $q=e^{-\pi t}$ throughout these formulae.

The Dedekind eta function is defined by
\begin{equation}
\eta(t) = q^{1/24} \prod_{n=1}^{\infty} (1-q^n)
\end{equation}
and the Jacobi theta function with general characteristic is defined as
\begin{equation}
\vartheta \left[\begin{array}{c} \alpha \\ \beta \end{array}\right] (z | t)= \sum_{n \in \mathbb{Z}} e^{-{(n+\alpha)}^2 \pi t/2} e^{2 \pi i (z+\beta)(n+ \alpha)} .
\end{equation}
The theta functions are manifestly invariant under $\alpha \rightarrow \alpha + \mathbb{Z}$ and also clearly obey
 \begin{equation}
\vartheta \left[\begin{array}{c} \alpha \\ \beta \end{array}\right] (z | t)= \vartheta \left[\begin{array}{c} \alpha \\ \beta +z \end{array}\right] (0 | t).
\end{equation}
A useful expansion valid for $\alpha \in (-\frac{1}{2} , \frac{1}{2}]$ is
\begin{equation}
\frac{\vartheta \left[\begin{array}{c} \alpha \\ \beta \end{array}\right] (0 | t)}{\eta(t)}= e^{2 \pi i \alpha \beta} q^{\frac{{\alpha}^2}{2}-\frac{1}{24}} \prod_{n=1}^{\infty}(1+e^{2 \pi i \beta}q^{n-\frac{1}{2} + \alpha})(1+e^{-2 \pi i \beta}q^{n-\frac{1}{2} - \alpha}) .
\end{equation}
For the four special theta functions we have
\begin{align}
{\vartheta}_1 (z|t) &\equiv \vartheta \left[\begin{array}{c} 1/2 \\ 1/2 \end{array}\right] (z | t) &=& \ 2 q^{1/8} \sin \pi z \prod_{n=1}^{\infty} (1-q^n)(1-e^{2\pi i z}q^n)(1-e^{-2 \pi i z} q^n), \\
{\vartheta}_2 (z|t) &\equiv \vartheta \left[\begin{array}{c} 1/2 \\ 0 \end{array}\right] (z | t) &=& \ 2 q^{1/8} \cos \pi z \prod_{n=1}^{\infty} (1-q^n)(1+e^{2\pi i z}q^n)(1+e^{-2 \pi i z} q^n), \\
{\vartheta}_3 (z|t) &\equiv \vartheta \left[\begin{array}{c} 0 \\ 0 \end{array}\right] (z | t) &=& \ \prod_{n=1}^{\infty} (1-q^n)(1+e^{2\pi i z}q^{n-\frac{1}{2}})(1+e^{-2 \pi i z} q^{n-\frac{1}{2}}), \\
{\vartheta}_4 (z|t) &\equiv \vartheta \left[\begin{array}{c} 0 \\ 1/2 \end{array}\right] (z | t) &=& \ \prod_{n=1}^{\infty} (1-q^n)(1-e^{2\pi i z}q^{n-\frac{1}{2}})(1-e^{-2 \pi i z} q^{n-\frac{1}{2}}).
\end{align}
These appear in the string partition function in the sum over spin structures. Derivatives w.r.t. $z$ give
\begin{align}
\vartheta_1(z) &= 2 \pi {\eta}^3 z + \mathcal{O}(z^3), \\
\vartheta_i(z) &= \vartheta_i(0) + \frac{z^2}{2} {\vartheta}_i^{\prime \prime}(0) + \mathcal{O}(z^4), \quad i=2,3,4
\end{align}
where we left the argument $t$ implicit.
In the course of expanding the vacuum string amplitude we arrive at expressions that can be simplified using a Riemann identity. In the $\mathcal{N}=1$ orbifold sector we can substitute
\begin{equation}
\sum_{\alpha, \beta = 0, 1/2} \eta_{\alpha \beta} \frac{{\vartheta}^{\prime \prime}\left[\begin{array}{c} \alpha \\ \beta \end{array}\right]}{\eta^3} \prod_{i=1}^{3} \frac{\vartheta\left[\begin{array}{c} \alpha \\ \beta + \theta_i \end{array}\right]}{\vartheta\left[\begin{array}{c} 1/2 \\ 1/2 + \theta_i \end{array}\right]} = -2 \pi \sum_{i=1}^3 \frac{{\vartheta}^{\prime}\left[\begin{array}{c} 1/2 \\ 1/2 - \theta_i \end{array}\right]}{\vartheta\left[\begin{array}{c} 1/2 \\ 1/2 - \theta_i \end{array}\right]},
\end{equation}
where $\eta_{\alpha \beta} = {(-1)}^{2(\alpha + \beta - 2 \alpha \beta)}$ and derivatives are w.r.t. $z$. When evaluating beta function coefficients it is useful to evaluate this for large $t$:
\begin{equation}
\lim_{t \rightarrow \infty} -2 \pi \sum_{i=1}^3 \frac{{\vartheta}^{\prime}\left[\begin{array}{c} 1/2 \\ 1/2 - \theta_i \end{array}\right]}{\vartheta\left[\begin{array}{c} 1/2 \\ 1/2 - \theta_i \end{array}\right]} = -2 \pi^2 \sum_i \frac{\cos \pi \theta_i}{\sin \pi \theta_i}.
\end{equation}
 We will rely on the following identity to simplify results in the $\mathcal{N}=2$ sector of the orbifold:
\begin{equation}
\sum_{\alpha, \beta = 0, 1/2} \eta_{\alpha \beta} {(-1)}^{2 \alpha} \frac{{\vartheta}^{\prime \prime}\left[\begin{array}{c} \alpha \\ \beta \end{array}\right]}{\eta^3} \frac{{\vartheta}\left[\begin{array}{c} \alpha \\ \beta \end{array}\right]}{\eta^3} \frac{\vartheta\left[\begin{array}{c} \alpha \\ \beta + \theta_1 \end{array}\right]}{\vartheta\left[\begin{array}{c} 1/2 \\ 1/2 + \theta_1 \end{array}\right]} \frac{\vartheta\left[\begin{array}{c} \alpha \\ \beta + \theta_2 \end{array}\right]}{\vartheta\left[\begin{array}{c} 1/2 \\ 1/2 + \theta_2 \end{array}\right]} = -4\pi^2,
\end{equation}
where $\theta_1 + \theta_2 =1$ mod 2.


\end{document}